\documentclass[journal]{IEEEtran}
\usepackage[T1]{fontenc}% optional T1 font encoding
\usepackage{cite}
\usepackage{amsmath,amssymb,amsfonts}
\usepackage[ruled,vlined]{algorithm2e}
\usepackage{graphicx}
\usepackage{textcomp}
\usepackage{xcolor}
\usepackage{booktabs}
\usepackage{threeparttable}
\usepackage{multirow}
\usepackage{diagbox}
\usepackage{url}
\usepackage{threeparttable}
\usepackage{array}
\usepackage{textcase}
\usepackage{etoolbox}
\makeatletter
\let\MYcaption\@makecaption
\makeatother

\usepackage[font=footnotesize]{subcaption}

\makeatletter
\let\@makecaption\MYcaption
\makeatother
\makeatletter
\def\UrlAlphabet{%
      \do\a\do\b\do\c\do\d\do\e\do\f\do\g\do\h\do\i\do\j%
      \do\k\do\l\do\m\do\n\do\o\do\p\do\q\do\r\do\s\do\t%
      \do\u\do\v\do\w\do\x\do\y\do\z\do\A\do\B\do\C\do\D%
      \do\E\do\F\do\G\do\H\do\I\do\J\do\K\do\L\do\M\do\N%
      \do\O\do\P\do\Q\do\R\do\S\do\T\do\U\do\V\do\W\do\X%
      \do\Y\do\Z}
\def\UrlDigits{\do\1\do\2\do\3\do\4\do\5\do\6\do\7\do\8\do\9\do\0}
\g@addto@macro{\UrlBreaks}{\UrlOrds}
\g@addto@macro{\UrlBreaks}{\UrlAlphabet}
\g@addto@macro{\UrlBreaks}{\UrlDigits}

\patchcmd{\@algocf@start}% <cmd>
  {-1.5em}% <search>
  {0pt}% <replace>
  {}{}% <success><failure>
\makeatother

\newcommand{\Times}[2]{${\text{#1}\times\text{#2}}$}
\newcommand{\SNR}[2]{${\text{SNR} #1 \text{#2}\;\text{dB}}$}
\def\post{\textit{a posteriori }}
\def\prior{\textit{a priori }}

\graphicspath{{figures/}}

\def\nt{N_{\rm t}}
\def\nr{N_{\rm r}}

\begin{document}
\setlength{\textfloatsep}{5pt}  
\setlength{\floatsep}{5pt}
\setlength\tabcolsep{3pt}%调表格列距
\ifdefined \GramaCheck
  \newcommand{\CheckRmv}[1]{}
  \newcommand{\figref}[1]{Figure 1}%
  \newcommand{\tabref}[1]{Table 1}%
  \newcommand{\secref}[1]{Section 1}
  \newcommand{\algref}[1]{Algorithm 1}
  \renewcommand{\eqref}[1]{Equation 1}
\else
  \newcommand{\CheckRmv}[1]{#1}
  \newcommand{\figref}[1]{Fig.~\ref{#1}}%
  \newcommand{\tabref}[1]{Table~\ref{#1}}%
  \newcommand{\secref}[1]{Sec.~\ref{#1}}
  \newcommand{\algref}[1]{Algorithm~\ref{#1}}
  \renewcommand{\eqref}[1]{(\ref{#1})}
\fi
\newtheorem{theorem}{Theorem}
\newtheorem{proposition}{Proposition}
\newtheorem{assumption}{Assumption}
\newtheorem{definition}{Definition}
\newtheorem{condition}{Condition}
\newtheorem{property}{Property}
\newtheorem{remark}{Remark}
\newtheorem{lemma}{Lemma}
\newtheorem{corollary}{Corollary}
\title{Gradient-Based Markov Chain Monte Carlo \\ for MIMO Detection}
\author{Xingyu~Zhou,
        Le~Liang,
        Jing~Zhang,
		    Chao-Kai~Wen,
        and~Shi~Jin% <-this % stops a space
\thanks{Xingyu Zhou, Jing Zhang, and Shi Jin are with the National Mobile
Communications Research Laboratory, Southeast University, Nanjing 210096, China
(e-mail: \protect \url{xy_zhou@seu.edu.cn}; jingzhang@seu.edu.cn; jinshi@seu.edu.cn).}% <-this % stops a space
\thanks{Le Liang is with the National Mobile
Communications Research Laboratory, Southeast University, Nanjing 210096, China, and also with Purple Mountain Laboratories, Nanjing 211111, China
(e-mail: lliang@seu.edu.cn).}% <-this % stops a space
\thanks{Chao-Kai Wen is with Institute of Communications Engineering,
National Sun Yat-sen University, Kaohsiung 80424, Taiwan
(e-mail: chaokai.wen@mail.nsysu.edu.tw).}% <-this % stops a space
}


\maketitle
\vspace{-2.0cm}
\begin{abstract}
Accurately detecting symbols transmitted over multiple-input multiple-output (MIMO) wireless channels is crucial in realizing the benefits of MIMO techniques.
However, optimal MIMO detection 
is associated with a complexity that grows exponentially with the MIMO dimensions and quickly becomes impractical.
Recently, 
stochastic sampling-based Bayesian inference techniques, such as Markov chain Monte Carlo (MCMC), have been combined with the gradient descent (GD) method to provide a promising framework for MIMO detection. 
In this work, we propose to efficiently approach optimal detection 
by exploring the discrete search space via MCMC random walk accelerated by {Nesterov's gradient} method.
Nesterov's GD guides MCMC to make efficient searches
without the computationally expensive matrix inversion and line search.  
Our proposed method operates using multiple GDs per random walk, achieving sufficient descent towards important regions of the search space before adding random perturbations, guaranteeing high sampling efficiency.
To provide augmented exploration, extra samples are derived through the trajectory of Nesterov's GD by simple operations, 
effectively supplementing the sample list for statistical inference and boosting the overall MIMO detection performance. 
Furthermore, we design an early stopping tactic to terminate unnecessary further searches,  remarkably reducing the complexity. 
Simulation results and complexity analysis reveal that the proposed method achieves {exceptional} performance in both uncoded and coded MIMO systems, adapts to realistic channel models, and scales well to large MIMO dimensions. 
\end{abstract}


% \fi
\IEEEpeerreviewmaketitle

%%%%%%%%%%%%%%%%%%%%%%%%%%%%%%%%%%%%%%%%%%%%%%%%%%%%%%%%%%
\section{Introduction}
% \vspace{-0.1cm}

\IEEEPARstart{I}n recent years, the multiple-input multiple-output (MIMO) technology has evolved towards the architecture of large-scale MIMO, deploying hundreds of antennas to achieve unprecedented spectral efficiency \cite{bjornson2017massive,ngoEnergySpectralEfficiency2013}.  
However, the benefits of MIMO depend fundamentally on efficient data detection at the receiver end. 
Among existing detection schemes, the optimal maximum \textit{a posteriori} (MAP) or maximum likelihood (ML) detection has a complexity exponentially increasing with the number of decision variables, which is impractical except for the most trivial cases \cite{yangFiftyYearsMIMO2015,albreemMassiveMIMODetection2019}. 
Meanwhile, linear detection, including the zero-forcing (ZF) and minimum mean square error (MMSE) methods, offers significant complexity reduction at the expense of suboptimal performance. 
Sphere decoding (SD)-based detectors are well-known for their near-optimal performance \cite{hochwaldAchievingNearcapacityMultipleantenna2003,guoAlgorithmImplementationKbest2006}. 
However, the tree search in SD still entails intensive complexity with the increase in modulation order and the number of antennas. 
Thus, {the realization of} an ideal trade-off between reception accuracy and efficiency has become the critical focus and challenge in MIMO detection scheme design.

Recently, machine learning has demonstrated its ability to address various challenges in wireless communications  \cite{eldar2022machine}. 
In particular, the Bayesian inference method is promising for dealing with the curse of dimensionality in MIMO detection. 
{The Bayesian inference resorts to approximation schemes to tackle the intractable posterior distribution of the transmitted symbols and can be broadly divided into two categories: deterministic and stochastic approaches  
\cite{bishopPatternRecognitionMachine}.} 

Deterministic approaches generally assume a specific factorization or parametric form (e.g., Gaussian) for analytical approximations to the posterior distribution.
Several related methods, such as the approximate message passing (AMP) \cite{donohoMessagepassingAlgorithmsCompressed2009} and the expectation propagation (EP) \cite{cespedesExpectationPropagationDetection2014}, have been widely exploited in designing MIMO detectors with outstanding performance and low complexity. 
However, deterministic approximations have an inherent disadvantage in that 
the exact results of the inference tasks can never be reached 
due to the various prerequisites and assumptions required by the analytical model \cite{heModelDrivenDeepLearning2020,kosasihGraphNeuralNetwork2022}. 
In contrast,
stochastic methods solve the inference problems based on numerical sampling and can model the complex posterior distribution with arbitrary levels of accuracy given sufficient computational resources.

Among stochastic sampling-based methods, the Markov chain Monte Carlo (MCMC) \cite{nealMCMCUsingHamiltonian2011} technique has received broad attention during the past several decades and has been widely employed in communication signal processing tasks \cite{farhang-boroujenyMarkovChainMonte2006,hedstromAchievingMAPPerformance2017,dattaNovelMonteCarloSamplingBasedReceiver2013,huangMCMCDecodingLDPC2020,wangOvertheAirAntennaArray2022}.  
Specifically, the MCMC-based MIMO detection is well recognized for its near-optimal performance \cite{farhang-boroujenyMarkovChainMonte2006,hedstromAchievingMAPPerformance2017,baiLargeScaleMIMODetection2016} and parallelizable architecture for efficient hardware implementation \cite{larawayImplementationMarkovChain2009}.
For MIMO detection, the MCMC method generates a list of samples
following the exact 
posterior distribution to perform statistical inference.
This sample list can be short, while capturing the statistics of the entire search space of the transmitted vectors \cite{hedstromAchievingMAPPerformance2017}. 
The Metropolis-Hastings (MH) algorithm \cite{hastings1970monte} is the most celebrated MCMC-type algorithm,  
which generally explores the search space 
in a simple random walk style to find important samples. 
However, the random walk exploration for high-dimensional search space requires an enormous number of steps before converging.  

The MCMC technique has recently been combined with the gradient descent (GD) method to improve the efficiency in finding important samples {and derive a gradient-based variant} \cite{maSamplingCanBe2019,wellingBayesianLearningStochastic,gowdaMetropolisHastingsRandomWalk2021}.  
{This combination bridges optimization and Monte Carlo sampling, which can be more powerful than conventional optimization algorithms, especially for nonconvex optimization settings \cite{maSamplingCanBe2019}.}  
For MIMO detection, {the authors of \cite{wuStochasticGradientLangevin2022}} have investigated the application of the stochastic gradient Langevin dynamics (SGLD) \cite{wellingBayesianLearningStochastic}, a gradient-based MCMC method, in building a stochastic MIMO detector. 
The SGLD-based detector achieves similar performance as the SD in medium-sized MIMO systems.
However, the design {in \cite{wuStochasticGradientLangevin2022}} is limited to the phase-shift keying modulation and is not applicable to the commonly used quadrature amplitude modulation (QAM). 
{In another study, the authors consider the discrete nature of the MIMO detection problem and derive an annealed Langevin dynamic to include the discrete prior of the transmitted symbols by injecting a sequence of noise with decreasing variance. 
This strategy makes possible the gradient computation and the sampling from the posterior distribution.} 
However, substantial running time is involved owing to the complicated
spectral domain analysis and the relatively large number of sampling iterations to achieve near-optimal performance \cite{zilbersteinAnnealedLangevinDynamics2022}.

Meanwhile, other authors propose to use a {preconditioned GD to accelerate the MCMC exploration} and derive a detector named MHGD, which  balances performance and complexity \cite{gowdaMetropolisHastingsRandomWalk2021}.
The MH random walk is moved in the preconditioned GD direction of the continuous-relaxed least squares (LS) surface to search for the ML estimate, 
preventing the bulk of ineffective searches in high-dimensional problems. 
However, the preconditioned GD component of MHGD belongs to Newton's method \cite{boyd2004convex} that requires high-order derivatives of the objective function and entails high-complexity matrix inversions and line search, precluding its application to large-scale MIMO systems. 
Moreover, the mini-batch property of the gradient methods is applied to MCMC algorithms, 
opening up the opportunity of extending the gradient-based MCMC to distributed MIMO detection \cite{wellingBayesianLearningStochastic,wuMinibatchMetropolisHastingsMCMC2019}.

{To enhance the performance of the MCMC detector, one can
strategically add extra samples from high probability regions in the search space to the sample list, referred to as sample augmentation (SA). 
Conventional schemes \cite{farhang-boroujenyMarkovChainMonte2006,hedstromAchievingMAPPerformance2017} simply increased the number of sampling iterations, 
causing a substantial rise in computational complexity and latency. 
An exception is the recent work presented in \cite{wangMarkovChainMonte2023}, wherein a crafted subset search complements the sample list to enhance soft decision accuracy. 
However, this scheme involves an additional search stage for the augmentation after the global search of MCMC, which introduces time overhead. 

Another critical issue in MCMC detector design is to determine when to terminate the sampling 
because a fixed number of iterations can lead to 
unnecessary searches that exert minimal impact on performance.
The early stopping (ES) technique from iterative search-based MIMO detectors is a potential solution to this issue, which terminates the search under certain conditions, conserving computational resources.
Existing ES strategies typically terminate the search when the residual norm from the current estimate falls below a predefined threshold \cite{dattaNovelMonteCarloSamplingBasedReceiver2013,dattaRandomRestartReactiveTabu2010}, 
or the residual norm remains unchanged for a certain number of iterations \cite{zhaoTabuSearchDetection2007,nguyenDeepLearningAidedTabu2020}.
However, depending solely on the residual norm can lead to unreliable final estimates and significant performance loss.}

In this work, we propose a {Nesterov's accelerated gradient (NAG)} \cite{nesterov1983method} aided MCMC sampling method, named NAG-MCMC, for MIMO detection.
We select Nesterov's GD because of its optimal convergence rate among the {first-order optimization methods using only the gradient evaluations}, thus making the proposed method highly efficient and scalable to high dimensions. 
Our proposed method follows the principle  
of gradient-based MCMC sampling and operates in a manner that accomplishes a satisfactory balance between Nesterov's GD and random walk, achieving high efficiency in finding the desired list of samples for detection. 
We also design additional strategies to enhance the performance and relieve the complexity of the proposed method. 
The contributions of this paper are summarized as follows:

\begin{itemize}
  
  \item \textit{Nesterov's accelerated gradient-based MCMC.}
  We utilize Nesterov's accelerated GD to guide the MCMC sampling while balancing the computational complexity. 
  Nesterov's GD provides efficient navigation over the continuous search space
  for the MCMC exploration, 
  and {MH} corrects the sample distribution via an acceptance test 
  and leads the algorithm to the important regions of the discrete search space. 
  We design a \textit{multiple GDs per random walk} strategy 
  to render sufficient descent before adding random perturbations, which improves the sampling efficiency. 
  The proposed NAG-MCMC achieves a comparable performance to the state-of-the-art MHGD while scaling well to large dimensions by avoiding the computations of matrix inversion and line search involved in the preconditioned GD.
  
  \item {\textit{Enhancing strategies for NAG-MCMC.} {We design the SA and ES strategies to improve the performance and reduce the computational complexity of NAG-MCMC. 
	These strategies are tailored to the distinctive structures of the proposed parallel NAG-MCMC detector, providing enhancements that diverge from conventional approaches.}  
	Specifically, SA utilizes simple operations to augment the sample list with additional important samples,  
  enhancing performance with minimal computational overhead.
  On the other hand, ES terminates the sampling when a large proportion of the samplers produce identical and sufficiently accurate best samples, effectively reducing the overall computational cost without perceivable performance degradation.}
  \item \textit{Comprehensive complexity analysis and  performance evaluation.} 
  We provide detailed computational complexity analysis to show that the NAG-MCMC detectors are well scalable to different MIMO dimensions with flexible complexity.
  We also conduct a wide range of numerical simulations to verify the potential of the proposed scheme. 
  Results demonstrate that the proposed scheme achieves {remarkable} performance
  in both uncoded and coded MIMO systems, adapts well to realistic channels, and outperforms a series of baselines in the trade-off between performance and complexity.
\end{itemize}

\textit{Notations:}
Boldface letters denote column vectors or matrices.
$\mathbf{0}_{M\times N}$ and $\mathbf{I}_N$ denote the $M\times N$-dimensional zero matrix and $N\times N$-dimensional identity matrix, respectively.
$(\cdot)^{T}$, $(\cdot)^{\rm H}$, and $(\cdot)^{-1}$ represent the transpose, conjugate transpose, and inverse, respectively.
$\|\cdot\|$ is $l_2$-norm, and $\|\cdot\|_F$ is Frobenius norm.
$\mathbb{E}[\cdot]$ represents the expectation of random variables.
$\mathcal{CN}(\mu,\sigma^2)$ denotes a complex-valued Gaussian distribution with mean $\mu$ and variance $\sigma^2$. $\mathcal{U}(a,b)$ indicates a uniform distribution between $[a,b]$. 

%%%%%%%%%%%%%%%%%%%%%%%%%%%%%%%%%%%%%%%%%%%%%%%%%%%%%%%%%%
\vspace{-0.2cm}
\section{System Model and Algorithm Review}
In this section, we first formulate the underlying problem for MIMO detection. We then briefly introduce the gradient-based MCMC sampling, which lays the foundation of our work.
\vspace{-0.5cm}
\subsection{System Model}

\CheckRmv{
  \begin{figure*}[t]
    \setlength{\abovecaptionskip}{-0.1cm}
	  \setlength{\belowcaptionskip}{-0.0cm}
    \centering
    \includegraphics[width=6.in]{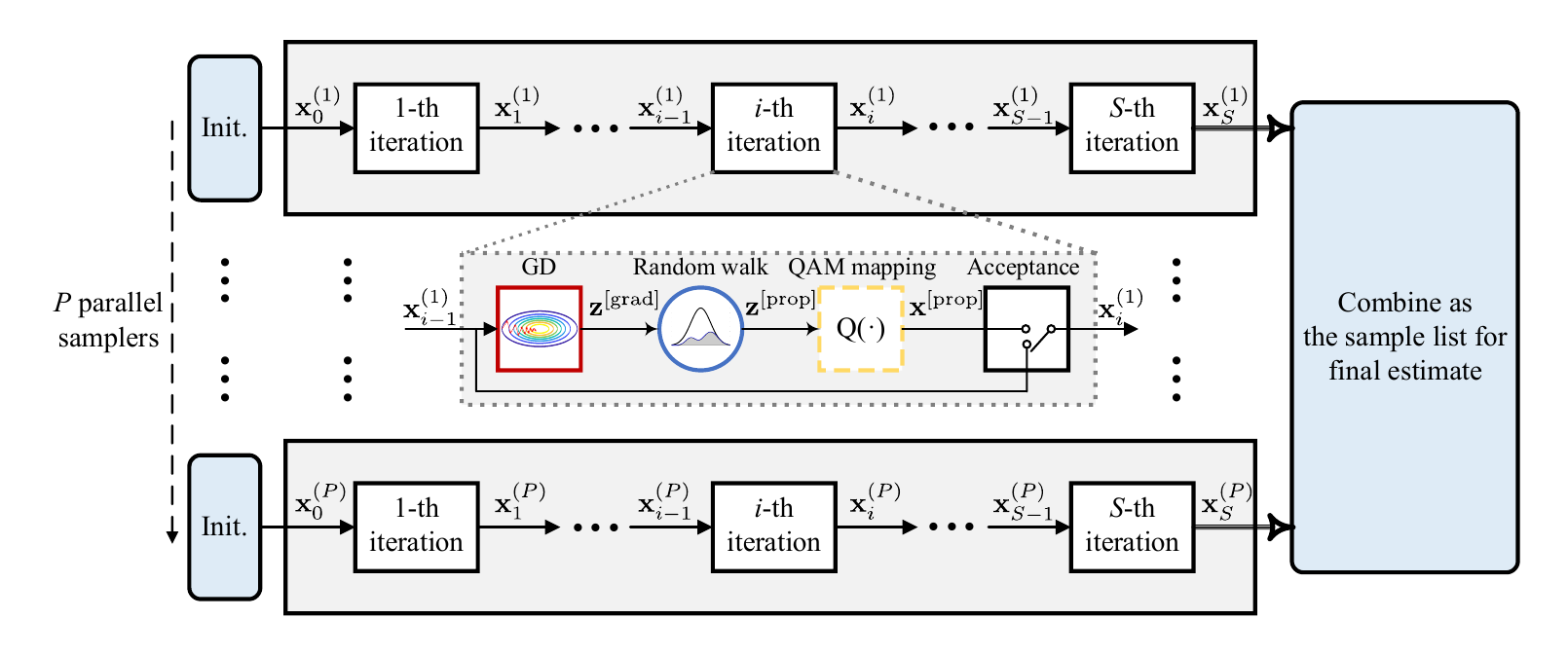}
    \caption{Block diagram of gradient-based MCMC sampling, where $P$ parallel samplers independently conduct $S$ sample draws to derive the sample list for the final estimate.
    Init.: Initialization.}
    \label{fig:sampler}
  \end{figure*} 
}

This paper considers a MIMO link with $\nt$ transmit antennas for sending independent data streams and $\nr$ antennas for receiving.
The input-output relationship of this spatial-multiplexing MIMO transmission is given by: 
\CheckRmv{
  \begin{equation}
    \mathbf{y}=\mathbf{Hx} + \mathbf{n},
    \label{eq:mimo_model}
  \end{equation}
}
where $\mathbf{y}\in \mathbb{C}^{\nr\times 1}$ is the channel observation at the receiver, $\mathbf{H}\in \mathbb{C}^{\nr \times \nt}$ is the flat-fading complex MIMO channel matrix, and 
$\mathbf{x}$ is the symbol vector, which contains $\nt$ QAM symbols mapped from the transmitted bits $\mathbf{b} \in \{\pm 1\}^{N_{\rm b}}$ of length $N_{\rm b}$. 
Letting $\mathcal{A}$ denote the QAM constellation set (normalized to obtain unit symbol energy) and $M$ be the cardinality of $\mathcal{A}$, we have $N_{\rm b}=\nt \log_2M$. 
$\mathbf{n}$ is the complex channel noise vector following $\mathcal{CN}(\mathbf{0}_{\nr\times 1},\sigma^2\mathbf{I}_{\nr})$ with $\sigma^2$ as the noise variance. The signal-to-noise ratio (SNR) is defined as $\text{SNR}=\mathbb{E}[\|\mathbf{Hx}\|^2] / \mathbb{E}[\|\mathbf{n}\|^2]$.

The MIMO detection problem is to estimate the transmitted vector $\mathbf{x}$ 
provided the triplet $\{\mathbf{y},\mathbf{H},\sigma^2\}$.
The optimal detector in the sense of minimum error probability aims at finding $\mathbf{x}$ that
maximizes its \post probability $p(\mathbf{x}|\mathbf{y})\propto p(\mathbf{x})p(\mathbf{y}|\mathbf{x})$, 
where $p(\mathbf{x})$  is the \prior distribution of $\mathbf{x}$, 
and $p(\mathbf{y}|\mathbf{x})$ is the conditional probability of the observation $\mathbf{y}$ (likelihood) given by: 
\CheckRmv{
  \begin{equation}
    p(\mathbf{y} |\mathbf{x})=\frac{1}{(\pi \sigma^2)^{\nr}} \exp \left(-\frac{1}{\sigma^2}\|\mathbf{y}-\mathbf{H x}\|^{2}\right).
    \label{eq:likelihood}
  \end{equation}
} 
Assuming that $\mathbf{x}$ is uniformly distributed over $\mathcal{A}^{\nt \times 1}$ and no other \prior information is available, 
the optimal MAP criterion is equivalent to the ML criterion to find the minimizer $\hat{\mathbf{x}}^{[\text{ML}]}$ of the cost function $f(\mathbf{x})=\frac{1}{2}\|\mathbf{y}-\mathbf{H x}\|^{2}$ in the discrete search space $\mathcal{A}^{\nt \times 1}$:  
\CheckRmv{
  \begin{equation}
    % \mathrm{ML}:\; 
    \hat{\mathbf{x}}^{[\text{ML}]}=\underset{\mathbf{x} \in \mathcal{A}^{{\nt \times 1}}}{\arg \min }\;f(\mathbf{x})=\frac{1}{2}\|\mathbf{y}-\mathbf{H x}\|^{2}, 
    \label{eq:ml}
  \end{equation}
}
We further define $\mathbf{r}=\mathbf{y}-\mathbf{H x}$ as the residual vector, and the cost function $f(\mathbf{x})$ can be equivalently represented by the squared residual norm $\|\mathbf{r}\|^2=\|\mathbf{y}-\mathbf{H x}\|^{2}$. 

In linear detection schemes, the constraint on the finite set of $\mathbf{x}$ can be relaxed to the continuous complex space $\mathbb{C}^{{\nt} \times 1}$ to derive the 
unconstrained LS solution of MIMO detection: 
\CheckRmv{
  \begin{equation}
    % \mathrm{LS}:\;
    \hat{\mathbf{x}}^{[\text{LS}]}=\underset{\mathbf{x} \in \mathbb{C}^{{\nt \times 1}}}{\arg \min }\;f(\mathbf{x})=\frac{1}{2}\|\mathbf{y}-\mathbf{H x}\|^{2}. 
    \label{eq:ls}
  \end{equation}
}
This relaxation transforms the non-deterministic polynomial-time hard (NP-hard) problem into a simple convex quadratic problem, remarkably reducing the complexity while delivering a highly-suboptimal solution, 
known as the ZF solution.

\subsection{Gradient-Based MCMC Sampling} \label{sec:mhgd}

The high computational complexity involved in finding the minimizer in \eqref{eq:ml} over the discrete search space of size $M^{\nt}$ 
makes the brute-force search for the ML infeasible for MIMO systems with high dimensions and high-order modulation.
To address this issue, MCMC has been increasingly used to derive a near-optimal solution \cite{farhang-boroujenyMarkovChainMonte2006,hedstromAchievingMAPPerformance2017}. 
MCMC methods can be classified into two categories: 
Gibbs sampling \cite{nealMCMCUsingHamiltonian2011} and the MH algorithm \cite{hastings1970monte}.
They both generate a sample sequence to construct a Markov chain that asymptotically converges to the target posterior distribution. 
The difference lies in that Gibbs sampling always accepts the proposals, 
whereas MH is a generalization that accepts or rejects the proposals based on an appropriate criterion. 
In addition, Gibbs sampling primarily operates on the bit or symbol level for MIMO detection \cite{farhang-boroujenyMarkovChainMonte2006,hedstromAchievingMAPPerformance2017,dattaNovelMonteCarloSamplingBasedReceiver2013}, updating only one variable (bit or symbol) at each step while the other variables remain unchanged.
In contrast, MH allows vectorizable operations by giving a proposal vector from the joint distribution of all the variables to be estimated, e.g., the $\nt \times 1$ transmitted symbols in MIMO detection. 
Therefore, MH can be parallelized for improved efficiency. 
The vector-level operations of MH also allow for the  
incorporation of gradients to expedite exploration over the search space.

In recent years, the combination of GD methods and the MCMC has shown great potential in solving nonconvex optimization problems \cite{maSamplingCanBe2019}. 
This approach leverages the 
GD to guide the search and exploits the random perturbation to escape local minima.
In this work, we focus on the gradient-based MCMC sampling framework, as presented in  
\figref{fig:sampler}.
The framework contains several parallel samplers independently performing $S$ sample draws to derive a sample list that approximates the target posterior. 
We use the superscript to distinguish samples drawn from different samplers in \figref{fig:sampler}, and we omit the superscript in the following description for simplicity.
The number of parallel samplers is denoted by $P$, 
and each sample draw 
is called a sampling iteration throughout the paper.
Each iteration includes the following steps: 
  \subsubsection{GD} {The relaxation in \eqref{eq:ls} allows the use of GD for optimization over the continuous complex space, which is first performed} to unveil the likely directions toward important regions of the discrete search space.
  Initialized with $\mathbf{z}_{0}=\mathbf{x}_{i-1}$, where $\mathbf{x}_{i-1}$ is the previous sample ($i$ is the sampling iteration index), the iterative optimization process of GD takes the basic form as:\footnote{We use the arguments $\mathbf{z}$ and $\mathbf{x}$ to represent the estimates in continuous and discrete search space, respectively.} 
  \CheckRmv{
    \begin{equation}
      \mathbf{z}_{t} = \mathbf{z}_{t-1} - \tau \nabla f(\mathbf{z}_{t-1}),\; t=1,\ldots,N_{\rm g},
      \label{eq:naive_gd}
    \end{equation}
  } 
  where $t$ is the GD iteration index, $\tau$ is the learning rate, and $\nabla f(\mathbf{z})$ denotes the gradient of the cost function: 
  \CheckRmv{
  \begin{equation}
    \nabla f(\mathbf{z})=-\mathbf{H}^{\rm H }(\mathbf{y} - \mathbf{Hz}).
    \label{eq:grad}
  \end{equation}
  } 
  The final GD estimate is ${\mathbf{z}}^{[\rm grad]} = \mathbf{z}_{N_{\rm g}}$ with $N_{\rm g}$ as a preset number of GD iterations.
  {The MCMC search can thus perform on the basis of the GD estimate, which significantly improves the quality of the proposal.} 
  Various GD methods can be used in the framework, such as the naive GD as shown in \eqref{eq:naive_gd}, Nesterov's GD \cite{nesterov1983method} introduced in \secref{sec:ngd}, or Newton's method \cite{boyd2004convex} that utilizes the second-order derivatives (i.e., Hessian) of the cost function to accelerate convergence. 
  Note that GD is an iterative optimization method, so there can be a sequential execution of \eqref{eq:naive_gd} 
  to improve the gradient estimate before proceeding to subsequent steps. 
   
  \subsubsection{Random Walk and QAM Mapping} 
  The MCMC exploration is conducted  
  in an MH random walk manner {after taking one or a few GD steps} to construct the proposal vector $\mathbf{x}^{[\rm prop]}$:
  \CheckRmv{
    \begin{align}
      \mathbf{z}^{[\rm prop]} &= \mathbf{z}^{[\rm grad]} + \mathbf{d}, \label{eq:rw1}\\
      \mathbf{x}^{[\rm prop]} &= Q(\mathbf{z}^{[\rm prop]}), 
      \label{eq:qam_map1}
    \end{align}
  }
  where $\mathbf{d}$ 
  is a Gaussian random vector, representing the perturbation injected by the random walk. 
  The random walk provides an opportunity for the sampler to escape local minima along the GD direction and explore the search space.
  However, to obtain a valid sample, the continuous update $\mathbf{z}^{[\rm prop]}$ should be discretized, which is accomplished by mapping each element of the vector to the closest QAM constellation point using the $Q(\cdot)$ function in \eqref{eq:qam_map1}. 
  This step, also known as the QAM mapping step, corrects the suboptimal LS solution and the inaccurate distribution 
  obtained by the GD over the continuous space, resulting in samples that {approximate} 
  the posterior distribution of the transmitted vector. 

  \subsubsection{Acceptance}  
  {According to the MH criterion \cite{hastings1970monte},
  the proposal $\mathbf{x}^{[\rm prop]}$ is accepted with the probability given by:}
  \CheckRmv{
  \begin{equation}
    \alpha  
    = \min \left\{1, \frac{\exp (-\|\mathbf{y}-\mathbf{H} \mathbf{x}^{[\rm prop]}\|^{2})}{\exp (-\|\mathbf{y}-\mathbf{H} \mathbf{x}_{i-1}\|^{2})}\right\},
    \label{eq:mh_pacc}
  \end{equation}}
  which can be viewed as the ratio between the likelihood evaluated at the proposal and that at the previous sample.

From \figref{fig:sampler}, 
we can see that the parallelized exploration of the search space provided by the independent $P$ samplers also benefits the gradient-based MCMC sampling. 
This significantly reduces the sampling latency and improves the diversity of the samples. 
Each sampler starts from an initial point that is either randomly chosen or transferred from another detector, such as MMSE. 
The sampler then runs for $S$ iterations to identify important samples over the discrete search space.
These samples are combined as a sample list $\mathcal{X}$, 
and hard decision $\hat{\mathbf{x}}$ can be generated by selecting the sample that minimizes the residual norm among $\mathcal{X}$:
\CheckRmv{
  \begin{equation}
    \hat{\mathbf{x}} = \underset{\mathbf{x} \in\mathcal{X}}{\arg \min }\;\|\mathbf{y}-\mathbf{Hx}\|^2.
    \label{eq:hard_output}
  \end{equation}
}

%%%%%%%%%%%%%%%%%%%%%%%%%%%%%%%%%%%%%%%%%%%%%%%%%%%%%%%%%%
\section{Nesterov's Accelerated Gradient-Based MCMC}
The details of the proposed NAG-MCMC are elaborated in this section.
First, we briefly review Nesterov's GD method \cite{nesterov1983method}. 
Then, we describe the parameter selection and  working principle of the NAG-MCMC method.

\vspace{-0.5cm}
\subsection{Nesterov's Accelerated Gradient Method} \label{sec:ngd} 
To begin, we review the heavy ball method \cite{polyak1964some}, which is closely related to the update rule used in Nesterov's accelerated GD.
In \eqref{eq:naive_gd}, the gradient $ - \nabla f(\mathbf{z})$ is the direction in which the cost function decreases the fastest.
Rather than simply stepping in this steepest direction, the heavy ball method 
incorporates some momentum from the previous update to accelerate and stabilize convergence. The revised \eqref{eq:naive_gd} is as follows:
\CheckRmv{
\begin{equation}
  \begin{aligned}
    {\mathbf{z}}_t &= {\mathbf{z}}_{t-1} - \tau \nabla f({\mathbf{z}}_{t-1}) + \rho ({\mathbf{z}}_{t-1} - {\mathbf{z}}_{t-2}) \\ 
    &={\mathbf{z}}_{t-1} - \tau \nabla f({\mathbf{z}}_{t-1}) + \rho \Delta\mathbf{z}_{t-1},
  \end{aligned}
\end{equation}
}
where 
$\Delta\mathbf{z}_{t-1} = {\mathbf{z}}_{t-1} - {\mathbf{z}}_{t-2}$ stands for the previous update ($\Delta\mathbf{z}_0=\mathbf{0}_{\nt\times 1}$), and $\rho \in [0,1]$ is the momentum factor dictating how much of the previous update 
is added to the current update. 
Nesterov's GD is a revision for the heavy ball method, which introduces an intermediate point in each iteration by adding the momentum $\rho\Delta\mathbf{z}_{t-1}$ to the previous state ${\mathbf{z}}_{t-1}$:
\CheckRmv{
  \begin{equation}
    \boldsymbol{p}_t = {\mathbf{z}}_{t-1} + \rho\Delta\mathbf{z}_{t-1},
    \label{eq:nesterov1}
  \end{equation}
}
then descends from the intermediate point $\boldsymbol{p}_t$:
\CheckRmv{
  \begin{equation}
    {\mathbf{z}}_t = \boldsymbol{p}_t - \tau \nabla f(\boldsymbol{p}_t). 
    \label{eq:nesterov2}
  \end{equation}
}

{For a convex cost function with an $L$-Lipschitz continuous gradient,} 
Nesterov demonstrated in \cite{nesterov1983method} that the update rule of \eqref{eq:nesterov1}-\eqref{eq:nesterov2} can achieve the optimal convergence rate of $\mathcal{O}(1/t^2)$ among the first-order gradient-based method 
by appropriately setting the learning rate $\tau$ and the momentum factor $\rho$.%setting to the reciprocal of the Lipschitz constant and {a carefully designed momentum factor sequence $\{\rho\}$}.
\footnote{The convergence rate means that in the $t$-th GD iteration, the gap $\epsilon = f({\mathbf{z}}_t)-f({\mathbf{z}}^{\ast})$ is on the order of $\mathcal{O}(1/t^2)$, where ${\mathbf{z}}^{\ast}$ minimizes $f$.} 
The most prominent advantage of Nesterov's method is its fast convergence using only the gradient of the cost function, 
without requiring higher-order derivatives that can be computationally expensive.
This property makes it a promising approach for addressing high-dimensional optimization problems.
{Additionally,  
Nesterov's GD has been shown to exhibit strong robustness to random perturbations during the update \cite{jinAcceleratedGradientDescent2018}, 
making it well-suited for integration with the random walk of MCMC in the gradient-based MCMC framework, as described in the following section.}

% \vspace{-0.5cm}
\subsection{The NAG-MCMC Algorithm} \label{sec:mhngd} 
The NAG-MCMC algorithm follows the gradient-based MCMC sampling framework by combining Nesterov's GD and MCMC sampling.
{In this subsection, we first discuss the detailed designs of these two components. 
Then, we introduce an approach based on \textit{multiple GDs per random walk} to coordinate these two components effectively.
Finally, we provide a summary of the proposed algorithm.} 
\subsubsection{{Design of Nesterov's GD}}  
{The following lemma shows 
that the cost function $f(\cdot)$ in \eqref{eq:ls} 
satisfies the constraints required by Nesterov's GD for achieving the convergence rate of $\mathcal{O}(1/t^2)$ \cite{nesterovIntroductoryLecturesConvex2004}.\footnote{Note that the optimality in the convergence rate of Nesterov's GD in the continuous-relaxed search space does not guarantee its optimality for the entire gradient-based MCMC tailored to the discrete search space of MIMO detection.} 
  \begin{lemma}
    \label{lemma2}
     The cost function $f(\cdot)$ in \eqref{eq:ls} is convex and has an  $L$-Lipschitz continuous gradient with the Lipschitz constant $L=\lambda_{\max}$,  that is, 
		\CheckRmv{
			\begin{equation}
				\|\nabla f(\mathbf{x}_1)-\nabla f(\mathbf{x}_2)\|\leq \lambda_{\max}\|\mathbf{x}_1-\mathbf{x}_2\|, \quad \forall\;\mathbf{x}_1, \mathbf{x}_2,
			\end{equation}
		}
    where $\lambda_{\max}$ is the maximum eigenvalue of $\mathbf{H}^{\rm H}\mathbf{H}$.
  \end{lemma}
  \begin{IEEEproof}
    The detailed proofs are left in the Appendix.
  \end{IEEEproof}
}
{Efficient parameter selection plays a crucial role in exploiting the strengths of Nesterov's GD. 
One key parameter to consider is the learning rate $\tau$, which can be selected based on \textit{Lemma}~\ref{lemma2}.
According to \cite{nesterovIntroductoryLecturesConvex2004}, for a cost function with an $L$-Lipschitz continuous gradient,}
any $\tau\leq 1/L$ can ensure the convergence of Nesterov's GD, and setting $\tau=1/L$ is a good choice to accelerate convergence. 
However, the eigenvalue decomposition for deriving $L=\lambda_{\max}$ is computationally expensive.
To avoid this tedious calculation and considering that $\lambda_{\max}$ is upper bounded by $\|\mathbf{H}^{\rm H}\mathbf{H}\|_{F}$, 
we set the learning rate as $\tau=1/\|\mathbf{H}^{\rm H}\mathbf{H}\|_{F}$ in our work and find negligible performance loss as compared with the choice of $\tau=1/\lambda_{\max}$. 

{Another parameter to consider is the momentum factor $\rho$, which influences acceleration. 
Following commonly accepted practices
\cite{ruderOverviewGradientDescent2017}, 
$\rho$ is set to a relatively large constant within $[0, 1]$. 
In our experiments, we have selected a value of $\rho=0.9$ to achieve sufficient acceleration.}

\subsubsection{{Design of MCMC Random Walk}}

The random walk is represented by the addition of a Gaussian random vector $\mathbf{d}$
as shown in \eqref{eq:rw1}, which can be further expressed as {$\mathbf{d} = \gamma \mathbf{M}_{\rm c}\mathbf{w}$, where $\gamma$ denotes the magnitude, i.e., the step size of the random walk, $\mathbf{M}_{\rm c} \in \mathbb{C}^{\nt \times \nt}$ specifies the covariance, and $\mathbf{w}\sim \mathcal{CN}(\mathbf{0}_{\nt\times 1}, \mathbf{I}_{\nt})$}.

{The general principle for designing the covariance and step size is to 
maintain an appropriate acceptance probability of the proposals while retaining the ability to escape local minima 
\cite{barbuHamiltonianLangevinMonte2020}.} 
{Regarding the covariance of $\mathbf{d}$,}  
we follow the choice of \cite{gowdaMetropolisHastingsRandomWalk2021} to set it 
proportional to $(\mathbf{H}^{\rm H} \mathbf{H})^{-1}$, approximating the spatial correlation between the noise of different data streams from the perspective of the transmitted vector $\mathbf{x}$. 
$\mathbf{M}_{\rm c}$ is set to the lower-triangular Cholesky factor of $(\mathbf{H}^{\rm H} \mathbf{H})^{-1}$ to construct this covariance. The channel gain of the Cholesky factor $\mathbf{M}_{\rm c}$ is then removed by performing a row normalization, 
 which retains the correlation while ensuring that $\mathbb{E}[(\mathbf{M}_{\rm c}\mathbf{w})(\mathbf{M}_{\rm c}\mathbf{w})^{\rm H}]=\mathbf{I}_{\nt}$ over random $\mathbf{H}$ and  
 $\mathbf{w}$.

Moreover, the step size of the random walk is given by
\CheckRmv{
  \begin{equation}
    \gamma = \max (d_{\rm qam},  \|\mathbf{r}\| / {\sqrt{\nr}}) \cdot \beta,
    \label{eq:stepsize}
  \end{equation}
}
where $d_{\rm qam}$ is half of the minimum Euclidean distance between any constellation points, and $\mathbf{r}$ is the residual vector after the QAM mapping step. 
This approach utilizes an adaptive step size that adjusts to the residual from the estimate, which is normalized by the dimension of the residual vector $\mathbf{r}$. 
When the current solution is far from optimal, large steps are taken, whereas a refined search is initiated when the residual is small enough. 
To avoid the stalling problem of MCMC and prevent the Markov chain from getting trapped around local minima, a minimum step size constraint $d_{\rm qam}$ is enforced \cite{farhang-boroujenyMarkovChainMonte2006,hedstromAchievingMAPPerformance2017}. 
{Furthermore, the coefficient $\beta$ in \eqref{eq:stepsize} should scale with the  dimension of the MIMO detection problem as $\beta \propto \nt^{-1/3}$ to maintain a reasonable acceptance probability of the proposal \cite{nealMCMCUsingHamiltonian2011}.}
We set the coefficient to $\beta = (\nt/8)^{-1/3}$ (e.g., $\beta=1$ for $\nt=8$ and $\beta=0.5$ for $\nt=64$) in our experiments and find this choice works well for different dimensions.

\subsubsection{{Multiple GDs per Random Walk}} 

\CheckRmv{
  \begin{figure}[t]
  \setlength{\abovecaptionskip}{-0.1cm}
  \centering
  \includegraphics[width=3.1in]{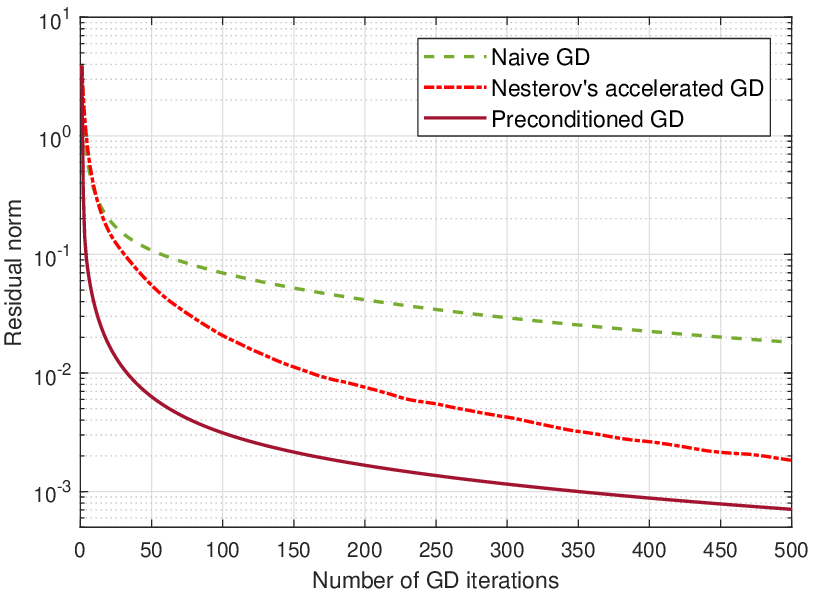}
  \caption{
   Residual norm from the estimated vector as a function of the number of GD iterations {in an \Times{8}{8} MIMO system with 16-QAM and \SNR{=}{20}. The naive GD uses the exact line search \cite{boyd2004convex} to determine the learning rate. The Nesterov's GD and preconditioned GD follow the configurations in \cite{nesterov1983method} and \cite{gowdaMetropolisHastingsRandomWalk2021}, respectively}.}  
  \label{fig:convergence_gd}
\end{figure}
}

We observe that in the framework described in \secref{sec:mhgd}, 
{how to coordinate the GD and the random walk, 
especially for the number of \textit{successive} GD iterations, i.e., $N_{\rm g}$, performed before the random walk step,}
is critical for the effectiveness of the algorithm for MIMO detection.
As a precursor to our approach, the MHGD detector proposed by \cite{gowdaMetropolisHastingsRandomWalk2021} inserts the random walk 
step after each GD iteration ($N_{\rm g}=1$).
This approach uses a preconditioned GD that belongs to Newton's method.
Preconditioned GD converges faster than Nesterov's GD because it utilizes high-order derivatives of the cost function, specifically the Hessian matrix, which is defined as $\nabla^{2} f(\mathbf{z})=\mathbf{H}^{\rm H }\mathbf{H}$.

The authors of \cite{gowdaMetropolisHastingsRandomWalk2021} utilize the inverse of a damped Hessian, $(\mathbf{H}^{\rm H}\mathbf{H}+\lambda\mathbf{I}_{\nt})^{-1}$, as the gradient preconditioner and line search for the optimal learning rate to accelerate the GD, where $\lambda>0$ is the damping parameter to construct a positive-definite preconditioner. 
\figref{fig:convergence_gd} shows the convergence rate comparison between the naive GD, Nesterov's GD, and preconditioned GD, {where the metric is defined as the residual norm from the GD estimate, $\|\mathbf{y}-\mathbf{Hz}_t\|$, with respect to the number of GD iterations.}
Despite a noticeable advantage of Nesterov's GD over naive GD, a non-negligible convergence rate gap can be found between Nesterov's GD and preconditioned GD in \figref{fig:convergence_gd}.
Thus, if Nesterov's GD is combined with random walk in a similar way as the MHGD in \cite{gowdaMetropolisHastingsRandomWalk2021}, the frequent random walk can disturb the descent directions, likely leading to poor performance {of the algorithm}.

\CheckRmv{
  \begin{figure}[t]
  \setlength{\abovecaptionskip}{-0.1cm}
  \centering
  \includegraphics[width=3.1in]{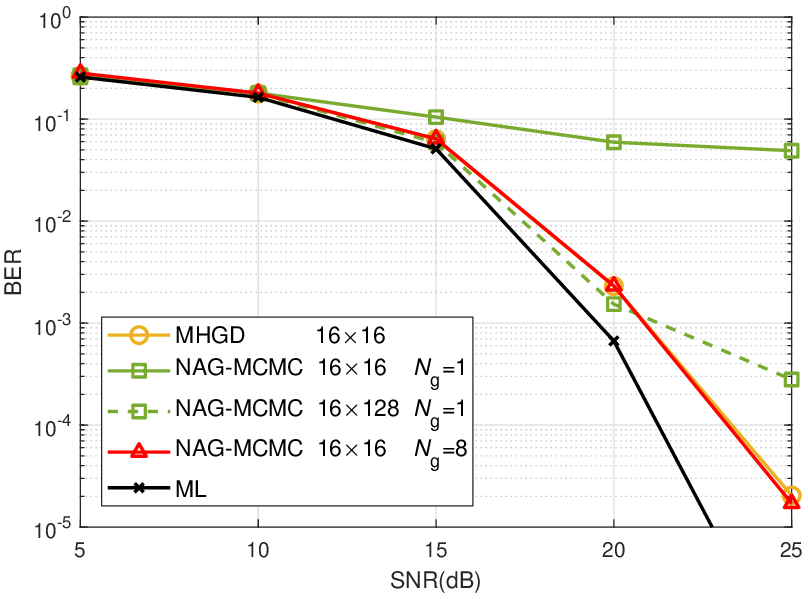}
  \caption{BER performance of NAG-MCMC with one GD or multiple GDs per random walk in an uncoded \Times{8}{8} MIMO system with 16-QAM under Rayleigh fading channels. The BER curves for MHGD and NAG-MCMC are marked by $P\times S$, representing the number of samplers and sampling iterations, respectively.}
  \label{fig:rationality}
\end{figure}
}

In NAG-MCMC, we propose a new approach referred to as \textit{multiple GDs per random walk},
where a series of Nesterov's GD iterations ($N_{\rm g}>1$) are performed before adding random perturbations. 
This approach has been largely unexplored in existing works \cite{maSamplingCanBe2019,wellingBayesianLearningStochastic,gowdaMetropolisHastingsRandomWalk2021,wuStochasticGradientLangevin2022,zilbersteinAnnealedLangevinDynamics2022}.
The successive GD iterations render a sufficient decrease in the cost function and lead the estimate to the regions closer to the ML solution, which significantly improves the sampling efficiency.

The results of following this line of thought 
are demonstrated in \figref{fig:rationality}, where we evaluate the bit error rate (BER) performance of different detectors 
under an uncoded \Times{8}{8} MIMO ($\nr\times \nt$) system with 16-QAM. 
The NAG-MCMC with \textit{eight} GDs per random walk ($N_{\rm g}=8$)
achieves virtually the same performance as MHGD, where both methods are equipped with the same number of samplers and sampling iterations ($P\times S=16\times 16$). 
{We remark that $N_{\rm g}$ 
is chosen to produce a similar effect achieved by one preconditioned GD iteration.} 
By contrast, the NAG-MCMC with one GD per random walk ($N_{\rm g}=1$) performs poorly. 
To further demonstrate the effectiveness of multiple GDs per random walk compared to one GD per random walk, we also showcase the performance of $S=128$ with $N_{\rm g}=1$ such that the total number of GDs performed, i.e., $N_{\rm g}S$, are the same for both cases. 
Note that this also leads to far more samples drawn 
and higher complexity for $N_{\rm g}=1$ when deciding the transmitted symbols according to \eqref{eq:hard_output}.  
From \figref{fig:rationality}, we conclude that adopting multiple GDs per random walk is strongly competitive, especially in the high SNR regime, where $N_{\rm g}=1$ suffers significant performance loss while having higher complexity.

\setlength{\algomargin}{0em} 
\SetAlCapHSkip{0em} 
\CheckRmv{
  \begin{algorithm}[!t]
    \SetKwInput{Compute}{Compute}
    \SetKwInput{Initialize}{Initialize}
    \SetKwInput{sone}{Step 1}
    \SetKwInput{stwo}{Step 2}
    \SetKwInput{sthree}{Step 3}
    \SetKwInput{sfour}{Step 4}
    \caption{NAG-MCMC}
    {
    \begingroup
    \KwIn{Channel observation $\bf y$, channel matrix $\bf H$, noise variance $\sigma^2$, {momentum factor $\rho$.}} 
    {\Compute{Learning rate $\tau$ for GD and 
    covariance $\mathbf{M}_{\rm c}\mathbf{M}_{\rm c}^{\rm H}$ for random walk.}}
    \ForPar{$p=1,\ldots,P$}{
    \Initialize{Initial estimate $\hat{\mathbf{z}}$ and $\mathbf{x}_0=Q(\hat{\mathbf{z}})$. Residual $\mathbf{r}_0 = \mathbf{y}-\mathbf{Hx}_0$.
    Momentum $\Delta{\mathbf{z}}_{1,0}=\mathbf{0}_{\nt\times 1}$. Random walk step size $\gamma=\max(d_{\rm qam}, \|\mathbf{r}_{0}\|/\sqrt{\nr})\cdot \beta$.} 
    \textbf{Core:}
    \For{$i=1,\ldots, S$}
    {
     \setlength\abovedisplayskip{0pt}
     \setlength\belowdisplayskip{0pt}
     \sone{Compute $N_{\rm g}$ Nesterov's GD iterations with ${\mathbf{z}}_{i,0}=\mathbf{x}_{i-1}$} 
     \For{$t=1,\cdots, N_{\rm g}$}{
      \setlength\abovedisplayskip{0pt}
      \setlength\belowdisplayskip{0pt}
      \begin{subequations}\label{eq:before_quan}%
        \begin{align}
        {\boldsymbol{p}}_{i,t}&={\mathbf{z}}_{i,t-1}+\rho \Delta {\mathbf{z}}_{i, t-1},\\
        {\mathbf{z}}_{i,t}&={\boldsymbol{p}}_{i,t} +  \tau\mathbf{H}^{\rm H}(\mathbf{y}-\mathbf{H}{\boldsymbol{p}}_{i,t}), \label{eq:gd}\\
        \Delta {\mathbf{z}}_{i,t} &={\mathbf{z}}_{i,t} - {\mathbf{z}}_{i,t-1}.% 
        \end{align}%
        \end{subequations}%
     }

     \textbf{Step 2:} 
     Derive the proposal with $\mathbf{z}^{[\rm grad]}=\mathbf{z}_{i,N_{\rm g}}$ and \\ 
     \quad\quad\quad\,\,$\mathbf{w} \sim \mathcal{CN}(\mathbf{0}_{\nt\times 1}, \mathbf{I}_{\nt})$
     \begin{subequations}\label{eq:alg_step2}
      \begin{align}
      \mathbf{z}^{[\rm prop]}&=\mathbf{z}^{[\rm grad]} + \gamma\mathbf{M}_{\rm c}\mathbf{w}, \label{eq:rw}\\
      \mathbf{x}^{[\rm prop]}&=Q(\mathbf{z}^{[\rm prop]}).  \label{eq:quan}
      \end{align}
      \end{subequations}
  
    \sthree{Acceptance test with $\mathbf{r}^{[\rm prop]} = \mathbf{y}-\mathbf{Hx}^{[\rm prop]}$, $u\sim\mathcal{U}(0,1)$, and
    \begin{equation}
      \alpha=\min\left\{1, \frac{\exp(-\|\mathbf{r}^{[\rm prop]}\|^2)}{\exp(-\|\mathbf{r}_{i-1}\|^2)}\right\}.
      \label{eq:alg_acc}
    \end{equation}}
    \If{$\alpha \geq u$}
    { \quad \quad \quad \quad 
      $\mathbf{x}_i=\mathbf{x}^{[\rm prop]}\;\; \text{and}\;\; \mathbf{r}_i=\mathbf{r}^{[\rm prop]}.$
    }
    \Else{\quad \quad \quad \quad 
      $\mathbf{x}_i=\mathbf{x}_{i-1}\;\; \text{and}\;\; \mathbf{r}_i=\mathbf{r}_{i-1}.$
    }
  
    \sfour{Update random walk step size with $\gamma=\max(d_{\rm qam}, \|\mathbf{r}_{i}\|/\sqrt{\nr})\cdot \beta$.} 
    }
    }
    \KwOut{$\hat{\mathbf{x}} = \underset{\mathbf{x} \in\mathcal{X}}{\arg \min }\;\|\mathbf{y}-\mathbf{Hx}\|^2$, where {$\mathcal{X}$ contains the samples $\{\mathbf{x}_i\}_{i=0}^{S}$ from the $P$ parallel samplers.}} 
    \endgroup
    }
  \label{alg:mhngd}
  \end{algorithm}
}

\subsubsection{{Algorithm Summary}}
Details of the NAG-MCMC MIMO detection method are summarized in \algref{alg:mhngd}. 
In each iteration of the core part for sample generation, $N_{\rm g}$ consecutive  iterations of Nesterov's GD following \eqref{eq:before_quan} are first performed over the previous sample $\mathbf{x}_{i-1}$. 
The subscript ${i,t}$ denotes the $t$-th GD iteration within the $i$-th sampling iteration, and $N_{\rm g}$ is a preset parameter 
to ensure an adequate descent from the previous point. 
After the GD, the proposal vector $\mathbf{x}^{[\rm prop]}$ is derived by the injection of random perturbation and QAM mapping in \eqref{eq:alg_step2} and accepted with probability $\alpha$ given in \eqref{eq:alg_acc} to generate the current sample.
{The procedure is repeated $S$ times, 
and $P$ parallel samplers are deployed to perform the method similar to \figref{fig:sampler}, complementing the desired sample list used to generate the decision.}
Finally, the hard output is provided as an exemplary usage of the sample list for MIMO detection in \algref{alg:mhngd}.

The core part of the algorithm is free from the gradient preconditioner and complicated line search required by Newton's method used in MHGD. 
Moreover, the required number of successive GDs, $N_{\rm g}$, for NAG-MCMC to achieve equivalent performance as the MHGD need not scale with the MIMO dimensions, which is demonstrated in the simulation results. 
These altogether make the proposed NAG-MCMC method promising for large-scale MIMO detection.

%%%%%%%%%%%%%%%%%%%%%%%%%%%%%%%%%%%%%%%%%%%%%%%%%%%%%%%%%%
\section{Enhanced NAG-MCMC}
We introduce several enhancements to further improve NAG-MCMC in this section, including the SA strategy for performance enhancement and the ES strategy for complexity reduction. 
We also extend the proposed method  
to support soft outputs in coded MIMO systems.

% \vspace{-0.5cm}
\subsection{Sample Augmentation for Performance Enhancement} \label{sec:sa}
When using the gradient-based MCMC algorithm for MIMO detection, the step of mapping the estimate to a discrete QAM constellation is critical for speeding up convergence to high probability regions in the discrete search space. 
However, 
no QAM mapping step is introduced between the consecutive GDs in NAG-MCMC, as seen in Step 1 of \algref{alg:mhngd}.
Thus, a large proportion of the search space is underexplored since no samples are derived over the consecutive GD procedure.

To compensate for this loss, we propose {the NAG-MCMC algorithm with SA}, deriving an augmented sample list that significantly enhances the performance. 
In particular, we introduce an additional QAM mapping operation after each descent in \eqref{eq:before_quan}, except for the $N_{\rm g}$-th descent after which the QAM mapping has already been inserted:
\CheckRmv{
  \begin{equation}
    \mathbf{x}^{[{\rm SA}]}_{(i-1)N_{\rm g}+t} =Q({\mathbf{z}}_{i,t}),\; i=1,\ldots,S,\; t=1,\ldots,N_{\rm g} - 1,
    \label{eq:sa}
  \end{equation}
}
where $\mathbf{x}^{[\rm SA]}_s$, with $s$ as the sample index, is the extra sample derived for the augmented sample list, and we define $\mathbf{x}^{[\rm SA]}_{iN_{\rm g}} = \mathbf{x}_i,i=0,\ldots ,S$, which is the sample derived in Step 2 and Step 3 of \algref{alg:mhngd}. 
{Note that \eqref{eq:sa} does not interrupt the original GD iterations but merely makes fuller use of previously discarded intermediate results with minimal complexity increase.}
Extra samples obtained from \eqref{eq:sa} are only used in the final decision stage, 
where the augmented sample lists $\mathcal{X}^{(p)}=\{\mathbf{x}^{[\mathrm{SA}]}_s\}_{s=0}^{N_{\rm g}S}$ from each of the $P$ samplers are concatenated to generate the final list $\mathcal{X}$. 

In summary, the change of the NAG-MCMC with SA compared to the initial NAG-MCMC method lies in the extra $(N_{\rm g} - 1)S$ samples derived from the trajectory of the successive Nesterov's GDs % NAG
by performing additional QAM mapping \eqref{eq:sa}, which complements the final list for decision and brings performance enhancement owing to the augmented exploration of the search space.

The extra computational overhead cost by SA 
is rather modest from the implementation perspective, only involving the QAM mapping and residual calculation for the extra samples when making final decisions with \eqref{eq:hard_output}.
The residual calculation is in the form of $\mathbf{r} =\mathbf{y}-\mathbf{H}{\mathbf{x}}$, where the elements of $\mathbf{x}$ are constellation points, i.e., $\mathbf{x}\in \mathcal{A}^{\nt \times 1}$.
{Considering that the QAM constellation has a series of unique magnitudes in their real or imaginary part, 
the matrix-vector multiplication $\mathbf{Hx}$ in the residual calculation can be substituted with the summation 
of the $\nt$ columns of $\mathbf{H}$ scaled by the QAM magnitudes. 
The scaled columns can be calculated in advance and stored for reuse in the iterations and \eqref{eq:hard_output} to prevent repeating multiplications.
Therefore, the overhead of the residual calculation for the extra samples can be reduced by a large extent.}
{The low-cost nature of the proposed SA is a notable advantage when compared to conventional techniques \cite{farhang-boroujenyMarkovChainMonte2006,hedstromAchievingMAPPerformance2017} that augment the sample list by increasing the number of iterations.
The latter approach often leads to a substantial increase in computational complexity and latency.}

\subsection{Early Stopping for Complexity Reduction} \label{sec:es}
In the gradient-based MCMC sampling framework, the preset number of sampling iterations can lead to unnecessary searches that have minimal impact on performance. 
Specifically, we observe that under many channel realizations, the samples drawn by the parallel NAG-MCMC-based {(including the NAG-MCMC with SA)} 
samplers become repetitious before reaching the preset number of iterations.
In fact, quite a few of these samples are precisely the ML estimate, indicating that high probability regions in the search space have been traversed by the parallel samplers, and the subsequent iterations hardly make contributions to performance improvement. 
Motivated by these observations, we design an ES strategy to avoid unnecessary searches and reduce the average number of iterations conducted by the algorithm.

At the end of each sampling iteration, we determine whether to stop the sampling by three steps.
First, from each of the $P$ samplers, the sample that minimizes the residual norm until the current iteration is collected, forming a set denoted as $\mathcal{X}^{[\min]}=\{\mathbf{x}^{[\min,(p)]}\}_{p=1}^{P}$. 
Second, the residual-norm-minimizer $\mathbf{x}^{[\min]}$ from the collected samples is selected:
\CheckRmv{
  \begin{equation}
    \mathbf{x}^{[\min]} = \underset{\mathbf{x} \in\mathcal{X}^{[\min]}}{\arg \min }\;\|\mathbf{y}-\mathbf{Hx}\|^2.
  \end{equation}
} 
The residual norm due to $\mathbf{x}^{[\min]}$ and the number of occurrences of $\mathbf{x}^{[\min]}$ in the collection $\mathcal{X}^{[\min]}$ are also computed, denoted by $l^{[\min]}=\|\mathbf{y} - \mathbf{Hx}^{[\min]}\|$ and $n^{[\min]}$, respectively.
{Finally, the following ES criterion is checked to determine whether the sampling shall be terminated:}
\CheckRmv{
  \begin{equation}
    n^{[\min]}> P/2 \quad \text{and} \quad l^{[\min]}<\eta \sqrt{\nr \sigma^2},  
    \label{eq:es}
  \end{equation}
}
{where the first half of the criterion in pertaining to the number of samplers that achieve the same minimizer.}
Note that the algorithm is generally configured with a relatively large number of samplers (e.g., 16 samplers) to improve parallelism.
Hence, when the minimizers of more than half of the samplers are the same, it is highly likely that the common minimizer $\mathbf{x}^{[\min]}$ is the optimal ML solution.  
The second half of the criterion \eqref{eq:es} is about whether the residual norm from $\mathbf{x}^{[\min]}$ is sufficiently small, where $\eta$ is a positive parameter to control this constraint and $\sqrt{\nr \sigma^2}$ is the expectation of the norm of the channel noise vector $\mathbf{n}$.
This constraint can effectively avoid terminating the algorithm when a large proportion of the samplers get trapped in the same local minimum.
In our experiment, we empirically set $\eta=1.5$ to effectively overcome unnecessary searches and meanwhile reduce the probability of inappropriate termination.

\CheckRmv{
  \begin{table}[t]
    \captionsetup[table]{skip=0pt}
    \centering
    \begin{threeparttable}
    \caption{Effects of the ES for an uncoded \Times{8}{8} MIMO  system with 16-QAM, ${\text{SNR} =25\;\mathrm{dB}}$, and Rayleigh fading channels %and $\eta=1.5$
    }
    % \vspace{-0.4cm}
    \begin{tabular}{l|lcccc} 
      \toprule
      {NAG-MCMC w/ SA}    & $S$  & 6             & 8             & 10            & 12             \\ 
      \cline{2-6}
      ($P=16, N_{\rm g}=8$)  & BER & 2.40e-4       & 6.00e-5       & 2.41e-5       & 1.00e-5        \\ 
      \midrule
      {NAG-MCMC w/ SA, ES} & $S_{\rm a}$  & \textbf{5.36} & \textbf{6.13} & \textbf{6.63} & \textbf{6.96}  \\ 
      \cline{2-6}
      ($P=16, N_{\rm g}=8$)  & BER & 2.45e-4       & 6.09e-5       & 2.45e-5       & 1.02e-5        \\
      \bottomrule
      \end{tabular}
    \label{tab:es}
    \begin{tablenotes}
      \footnotesize
      \item[]Note: {The results are averaged over 312,500 channel realizations.}
    \end{tablenotes}
  \end{threeparttable}
  \end{table}
}

The effectiveness of the proposed ES strategy is shown in \tabref{tab:es}, where an \Times{8}{8} MIMO system with 16-QAM and \SNR{=}{25} is considered. 
{The table presents the BER performance of NAG-MCMC with both SA and ES (NAG-MCMC w/ SA, ES) versus NAG-MCMC with SA but no ES adopted (NAG-MCMC w/ SA) under different numbers of sampling iterations $S$.}
In the table, $S_{\rm a}$ is the average number of iterations performed by the NAG-MCMC with SA and ES given $S$ in the corresponding column, where the results are averaged over 312,500 channel realizations.

{The comparison in \tabref{tab:es} demonstrates that NAG-MCMC with both SA and ES has negligible performance loss compared to NAG-MCMC with SA but no ES adopted while significantly reducing the average number of sampling iterations.}
Note that with the increase of $S$, the BER performance of NAG-MCMC with SA improves at the expense of additional waste of iterations for certain channel realizations where the ML solution can be found with a smaller number of iterations than $S$. 
In this situation, the advantage of using ES is more prominent. 
Specifically, when $S=12$, $S_{\rm a}$ is approximately 6.96, which is nearly reduced by half. 

{\begin{remark}
While the motivation for using ES is similar to its applications in other search-based MIMO detectors \cite{nguyenDeepLearningAidedTabu2020,dattaRandomRestartReactiveTabu2010,zhaoTabuSearchDetection2007}, 
the mechanism employed in NAG-MCMC is distinct from existing solutions.
Specifically, we have taken into account the characteristics of parallel sampling when designing the ES criterion, as indicated by the first criterion in \eqref{eq:es}, which has not been explored in previous works \cite{nguyenDeepLearningAidedTabu2020,dattaRandomRestartReactiveTabu2010,zhaoTabuSearchDetection2007}.
This particular design enhances the reliability of the solution $\mathbf{x}^{[\min]}$ when the search is terminated early.{\footnote{{A similar strategy can be found in \cite{dattaNovelMonteCarloSamplingBasedReceiver2013}, where the number of repetitions of the best sample is incorporated within the ES criterion. However, that strategy is designed for a sequential search process, in contrast to the parallel sampling used in NAG-MCMC, influencing decisions on whether to terminate or restart the search.}}} 
Through experiments, we have observed that the use of \eqref{eq:es} yields superior results compared to the conventional approach that solely relies on a residual norm threshold 
($l^{[\min]}<\eta \sqrt{\nr \sigma^2}$) \cite{dattaRandomRestartReactiveTabu2010}.
\end{remark}}

\subsection{Applied to Coded MIMO Systems} \label{sec:coded} 
In this subsection, we investigate the application of the proposed schemes to coded MIMO systems, where the NAG-MCMC detector and the enhanced versions use the collected sample list to generate soft outputs.
Most current wireless communication systems utilize error correction coding schemes to improve performance, and the MIMO detector outputs soft reliability information, typically in the form of log-likelihood ratios (LLRs).
These LLRs are further processed by the channel decoder to derive a more reliable decoding result. 
The \post LLR for the $k$-th entry $b_k$ of the transmitted bit vector $\mathbf{b}$ is defined as:
\CheckRmv{
  \begin{equation}
  L_{k} = \log \frac{p\left(b_{k}=+1 |\mathbf{y}\right)}{p\left(b_{k}=-1 | \mathbf{y}\right)}
 =\log \frac{\sum_{\mathbf{x} \in \mathcal{A}_{k+}^{\nt \times 1}} p(\mathbf{y} | \mathbf{x})p(\mathbf{x})}{\sum_{\mathbf{x} \in\mathcal{A}_{k-}^{\nt \times 1}} p(\mathbf{y} | \mathbf{x})p(\mathbf{x})},
  \label{eq:llr}
\end{equation} 
}
where $\mathcal{A}_{k+}^{\nt \times 1}$ and $\mathcal{A}_{k-}^{\nt \times 1}$ are the subsets of $\mathcal{A}^{\nt \times 1}$ for the symbol vector $\mathbf{x}$ mapped from $\mathbf{b}$ with $b_k$ corresponding to $+1$ and $-1$, respectively. 
The NAG-MCMC-based detectors can approximate {$L_k$ in \eqref{eq:llr}} 
on the basis of the collected sample list $\mathcal{X}$ from the parallel samplers, which is substantially smaller than the set $\mathcal{A}^{\nt \times 1}$ of size $M^{\nt}$, 
and the max-log approximation \cite{robertson1995comparison} is typically used for simplification:
\CheckRmv{
  \begin{equation}
    \begin{aligned}
      L_{k}& \approx \min _{\mathbf{x} \in \mathcal{X}_{k-}}\left\{\frac{1}{\sigma^{2}}\|\mathbf{y}-\mathbf{H} \mathbf{x}\|^{2} -\frac{1}{2}\mathbf{b}^{T}\mathbf{L}_{{\rm a}}\right\}\\
      &-\min _{\mathbf{x} \in \mathcal{X}_{k+}}\left\{\frac{1}{\sigma^{2}}\|\mathbf{y}-\mathbf{H} \mathbf{x}\|^{2} -\frac{1}{2}\mathbf{b}^{T}\mathbf{L}_{{\rm a}}\right\},
    \label{eq:max_log}
    \end{aligned}
  \end{equation}
}
where the subsets $\mathcal{A}_{k+}^{\nt \times 1}$ and $\mathcal{A}_{k-}^{\nt \times 1}$  in \eqref{eq:llr} are replaced by $\mathcal{X}_{k+} = \mathcal{X} \cap \mathcal{A}_{k+}^{\nt \times 1}$ and $\mathcal{X}_{k-}=\mathcal{X} \cap \mathcal{A}_{k-}^{\nt \times 1}$, respectively, 
{and $\mathbf{L}_{{\rm a}}$ represents the vector of \prior information 
on $\mathbf{b}$ and $\mathbf{L}_{{\rm a}}=\mathbf{0}_{N_{\rm b}\times 1}$ when no \prior information is available.  
The NAG-MCMC-based detectors can also support iterative detection and decoding \cite{hochwaldAchievingNearcapacityMultipleantenna2003} to enhance performance, with $\mathbf{L}_{{\rm a}}$ from the channel decoder and calculating extrinsic soft outputs by subtracting the priors from \eqref{eq:max_log}, as shown in \cite{farhang-boroujenyMarkovChainMonte2006,hedstromAchievingMAPPerformance2017}.}

Note that the LLR calculation in \eqref{eq:max_log} requires that $\mathcal{X}_{k+} \neq \emptyset$ and $\mathcal{X}_{k-} \neq \emptyset$. 
In other words, at least one sample with $b_k=1$ and $b_k=-1$ should be found. 
This condition cannot be guaranteed considering that the sample list $\mathcal{X}$ is incomplete compared to the entire search space $\mathcal{A}^{\nt \times 1}$. 
{However, the lack of samples happens only when $b_k$ takes $+1$ (or $-1$) with a low probability because the sample list $\mathcal{X}$ captures the statistics of $\mathcal{A}^{\nt \times 1}$.}
Therefore, high confidence can be given to this $k$-th bit, and the LLR can be directly set to a saturated value \cite{hedstrom2021capacity}.
In our experiment, we set the saturated level to $\pm 10/\sigma^2$ to achieve accurate and stable decoding.

\newcolumntype{M}[1]{>{\centering\arraybackslash}m{#1}}
\CheckRmv{
  \begin{table*}[t]
    \captionsetup[table]{skip=0pt}
    \centering
    \begin{threeparttable}
    \caption{Computational Complexity Comparsion}
    % \vspace{-0.2cm}
    \begin{tabular}{lM{28em}} 
    \toprule
    \textbf{Algorithms} & \textbf{Number of complex multiplications} \\ 
    \hline\hline
    MMSE   &  $3N^3 + 2N^2 + 2N$      \\
    \hline
    EP    &  $\big(2N^3 + N^2 + (2M+13)N\big) T + N^3 + 2N^2 + N$ \\
    \hline
    MHGD  &  $\big(6N^2 + (M+5)N\big)P S + 23/3N^3 + (M+2)N^2 +7N$    \\
    \hline
    NAG-MCMC & $\big((N_{\rm g}+1)N^2 + (M+2N_{\rm g}+2)N\big)P S +  (M+1)N P + 11/3N^3 + (M+2)N^2 + 2N$   \\
    \hline
    NAG-MCMC w/ SA, ES &  $\big((N_{\rm g}+1)N^2 + ((M+3)N_{\rm g}+1)N\big)PS_{\rm a} + (M+1)N P + 11/3N^3 + (M+2)N^2 +2N$   \\
    \bottomrule
    % \multicolumn{2}{l}{%
    %   \begin{minipage}{40em}~\\
    %     \footnotesize $N$: Num. of antennas; $M$: Modulation order; $T$: Num. of EP iterations; $P$: Num. of samplers; $S$: Num. of samples; $N_{\rm g}$: Num. of inner GDs; $N_{\rm nz}$: Average num. of nonzero elements in $\Delta \mathbf{x}$.        
    %   \end{minipage}
    % }
    \end{tabular}
    \label{tab:complexity}
    % \begin{tablenotes}[para,flushleft]
    %   \footnotesize
    %   \item[]$N$: Number of antennas; $M$: Cardinality of the constellation; $T$: Number of EP iterations; $P$: Number of samplers; $S$: Number of sampling iterations; $N_{\rm g}$: Number of successive Nesterov's GD iterations.
    % \end{tablenotes}
  \end{threeparttable}
  \end{table*}
}
%%%%%%%%%%%%%%%%%%%%%%%%%%%%%%%%%%%%%%%%%%%%%%%%%%%%%%%%%%
\section{Complexity Analysis}
The key advantage of the proposed NAG-MCMC and its enhanced versions lies in their scalability to large-dimensional systems with low complexity. 
We analyze the complexity of the proposed methods to demonstrate this advantage in this section.
In the complexity analysis, we assume that $\nt = \nr = N$ for a simplified notation of the complexity expressions.

\vspace{-0.2cm}
\subsection{Complexity Expressions}

\tabref{tab:complexity} compares the computational complexity of various MIMO detectors, based on the number of complex multiplications needed to detect a single symbol vector.
{We use this metric because it dominates the number of required floating-point operations compared to other operations, such as additions, divisions, and square roots.}  
The inversion of a symmetric matrix involved in the compared schemes is calculated by Cholesky decomposition, which costs $2(N^3 + N)$ complex multiplications  \cite{maQRDecompositionBasedMatrix2011}. 
The EP detector \cite{cespedesExpectationPropagationDetection2014} involves $T$ iterations, and each iteration contains a matrix inversion.
The MHGD algorithm  \cite{gowdaMetropolisHastingsRandomWalk2021} entails three matrix inversions, including the MMSE initializer, the gradient preconditioner for GD acceleration, and the random walk covariance.
In contrast, the proposed NAG-MCMC requires only one matrix inversion, which is used for the computation of the random walk covariance.
Note that the calculation of $\mathbf{H}^{\rm H}\mathbf{H}$ and the Cholesky factor $\mathbf{M}_{\rm c}$ also costs $N^3$ and $2/3 N^3$ multiplications \cite{trefethen1997numerical}, respectively{, hence accumulated as the $11/3 N^3$ expression in \tabref{tab:complexity}, reduced by $4N^3$ compared to the counterpart in MHGD}.

The core part of NAG-MCMC is dominated by the matrix-vector multiplications, $N_{\rm g}$ times for Nesterov's GDs 
in \eqref{eq:gd} and one for the random walk step in \eqref{eq:rw}
as shown in \algref{alg:mhngd}, costing $(N_{\rm g}+1)N^2$ complex multiplications in total.\footnote{The complex multiplications cost by the residual calculation $\mathbf{r}=\mathbf{y}-\mathbf{Hx}$ are moved to the preprocessing stage for calculating the scaled columns of $\mathbf{H}$ and the cost is $MN^2$ in total.}
The remaining computations are from the QAM mapping, and the cost is modest. 
The enhanced NAG-MCMC with SA and ES has a small complexity increase due to additional residual calculation and QAM mapping for the augmented samples.  
The complexity caused by the additional operations is modest considering the unified preprocessing for the residual calculation and the simple implementation of QAM mapping.
Furthermore, the ES strategy can reduce the required number of iterations from the preset $S$ to the substantially smaller $S_{\rm a}$ as shown in \secref{sec:es}.
Thus, the NAG-MCMC with SA and ES would achieve a desirable trade-off between complexity and performance.

\subsection{Complexity Scaling Behavior}

We analyze the computational complexity of the proposed algorithms, focusing on how they scale with the number of antennas and compare them with the baselines in \figref{fig:complexity_scaling}.
The parameter configurations of the compared schemes are chosen for comparable performance, which is demonstrated in \secref{sec:simulation}, except for the MMSE which cannot reach the considered performance level. 
Specifically, the EP method runs for $T=10$ iterations.
{The MHGD baseline, NAG-MCMC, and NAG-MCMC with SA and ES are all equipped with $P=16$ samplers.} 
The preset number of iterations is $S=8$. 
The average number of iterations in NAG-MCMC with SA and ES  
is $S_{\rm a}\approx 5.0$, which ensures an equivalent performance to the baselines. 
The number of successive Nesterov's GD iterations in NAG-MCMC and NAG-MCMC with SA and ES is $N_{\rm g} = 8$.

\CheckRmv{
  \begin{figure}[t]
  \setlength{\abovecaptionskip}{-0.1cm}
  \centering
  \includegraphics[width=3.1in]{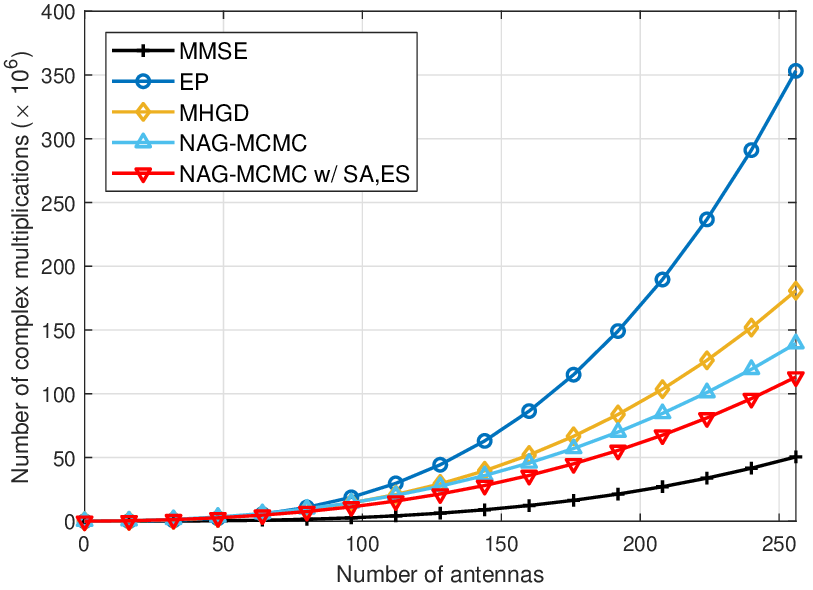}
  \caption{Complexity scalinig behavior with respect to the number of antennas with $\nt=\nr\in [0,256]$ and 16-QAM modulation.}
  \label{fig:complexity_scaling}
  \end{figure}
}

\figref{fig:complexity_scaling} shows that the EP method has a complexity advantage in medium-sized MIMO systems, e.g., $N=8$, as noted in \cite{gowdaMetropolisHastingsRandomWalk2021}.
However, the proposed NAG-MCMC with SA and ES becomes less complex than the EP and MHGD baselines when the system dimension increases.
Specifically, the superiority appears  
when the number of antennas is around 50--60 and becomes more pronounced when the number of antennas is more than 150. 
Moreover, the cost of NAG-MCMC with SA and ES is only about 2.24 times that of the MMSE linear receiver at the largest system dimension considered ($N=256$).  
This superiority is due to the reduction of matrix inversions 
and the high searching efficiency of the proposed method. 
Therefore, the NAG-MCMC with SA and ES scales well to high-dimensional systems and is the most promising approach for balancing performance and complexity among the compared schemes.  

{\begin{remark}
  It is noteworthy 
  that the computational complexity of the NAG-MCMC-based detectors presented in this section is tailored for near-ML performance in 
  general MIMO settings. 
  However, for specific MIMO configurations, 
  further complexity reductions can be attained through various approximations.
  For example, in massive MIMO systems with $\nr\gg \nt$, experimental observations suggest that a diagonal approximation can be applied when calculating $\mathbf{H}^{\rm H}\mathbf{H}$. 
  Additionally, an identity matrix can be employed as the random walk covariance instead of $(\mathbf{H}^{\rm H}\mathbf{H})^{-1}$.
  These approximations reduce the complexity of the proposed detector from being cubic to quadratic in MIMO dimensions, while incurring virtually no performance loss.
\end{remark}
}

%%%%%%%%%%%%%%%%%%%%%%%%%%%%%%%%%%%%%%%%%%%%%%%%%%%%%%%%%%
\section{Simulation Results} \label{sec:simulation}
The proposed NAG-MCMC and enhanced versions are evaluated by numerical simulations in this section. 
First, the parameter settings are described. 
Second, the performance of the proposed schemes in uncoded MIMO systems is presented.
Finally, the application of the proposed methods to coded MIMO systems is evaluated.

% \CheckRmv{
%   \begin{figure}[t]
%     % \setlength{\abovecaptionskip}{-0.1cm}
%     \centering
%     \subfigure[\Times{8}{8} MIMO, 16-QAM, \SNR{=}{25}, $P=16$]{
%       \includegraphics[width=3.1in]{convergence_8MIMO_16QAM_25dB.eps}
%       \label{fig:convergence_8mimo}
%     }
%     \subfigure[\Times{16}{16} MIMO, 64-QAM, \SNR{=}{30}, $P=32$]{
%       \includegraphics[width=3.1in]{convergence_16MIMO_64QAM_30dB.eps}
%       \label{fig:convergence_16mimo}
%     } 
%     \vspace{-0.8cm}
%     \caption{BER convergence property with respect to the number of sampling {iterations} in 
%     %The number of parallel samplers is 
%     % $P=16$ for the 16-QAM case and $P=32$ for the 64-QAM case, 
%     Rayleigh fading channels.}
    
%     \label{fig:convergence}
%   \end{figure}
% }

\vspace{-0.3cm}
\subsection{Parameter Settings} \label{sec:parameter}
The Rayleigh fading and 3rd generation partnership project (3GPP) channel models \cite{3gpp38901} are adopted in the simulations, and perfect channel state information (CSI)  
is assumed to be available {unless noted otherwise}. 
The Rayleigh fading channels have independent Gaussian fading coefficients with zero mean and variance $1/\nr$.
Both uncoded and coded MIMO systems are investigated. 

For uncoded MIMO systems, the maximum number of transmitted bits is set to $10^8$ to evaluate the BER performance. 
The baselines include the conventional MMSE method, the machine learning-based EP iterative detector \cite{cespedesExpectationPropagationDetection2014}, the state-of-the-art MHGD detector \cite{gowdaMetropolisHastingsRandomWalk2021}, 
{the near-optimal $K$-best SD detector \cite{guoAlgorithmImplementationKbest2006},} 
and the optimal ML detector.   
EP is set to 10 iterations and the damping factor is 0.2 to achieve the performance limit of this method corresponding to the convergence analysis in \cite{cespedesExpectationPropagationDetection2014}. 
{The performance does not improve beyond this limit since EP is a deterministic approximation scheme.}
The parameters for simulating the MHGD are the same as those in \cite{gowdaMetropolisHastingsRandomWalk2021} to provide a near-ML baseline.  
Unless noted otherwise, the MHGD is initialized by the MMSE initializer according to \cite{gowdaMetropolisHastingsRandomWalk2021},  
and the proposed NAG-MCMC-type algorithms 
randomly select the initial estimate vector from $\mathcal{A}^{\nt \times 1}$.
The MHGD and the proposed methods produce hard outputs by selecting the best sample that minimizes the residual norm among the generated sample list according to \eqref{eq:hard_output}.

% \CheckRmv{
%   \begin{figure}[t]
%     \centering
%     \subfigure[16-QAM]{
%       \includegraphics[width=3.1in]{BER_8MIMO_16QAM_1.eps}
%       \label{fig:ber_8mimo_16qam_1}
%     }
%     \subfigure[64-QAM]{
%       \includegraphics[width=3.1in]{BER_8MIMO_64QAM_1.eps}
%       \label{fig:ber_8mimo_64qam_1}
%     } 
%     \vspace{-0.8cm}
%     \caption{BER performance comparsion between NAG-MCMC and its enhanced versions in an \Times{8}{8} MIMO system with 16-QAM/64-QAM modulation and Rayleigh fading channels.
%     % for the \Times{8}{8} MIMO case with 16-QAM and 64-QAM modulation. %$S_{\max}=5$ for NAG-MCMC with SA and ES when \SNR{\leq}{15}.
%     } % $S_{\max}=5$ for 16-QAM and 64-QAM
%     \label{fig:ber_8mimo_1}
%   \end{figure}
% }

For coded MIMO systems, 
low-density parity-check (LDPC) codes with a code rate of 3/4 and block length of 1944 bits from the IEEE 802.11n standard are used as the error correction codes. 
Belief propagation with 10 inner iterations is used for channel decoding, which is the same for all compared detectors.
The block error rate (BLER) is selected as the performance metric of coded MIMO systems and is evaluated until the number of block errors exceeds 10 or a maximum number of $10^5$ blocks are transmitted. 
For the baselines, EP is revised according to \cite{cespedesProbabilisticMIMOSymbol2018} for the coded systems.
The MHGD 
and the proposed methods generate soft outputs in terms of LLRs following the means described in \secref{sec:coded}.
Meanwhile, a $K$-best SD algorithm \cite{guoAlgorithmImplementationKbest2006} is simulated to provide a near-ML baseline.

\CheckRmv{
  \begin{figure}[t]
    \captionsetup[subfigure]{aboveskip=-0.5pt,belowskip=-0.5pt}
    \centering
    \begin{subfigure}{3.1in}
      \includegraphics[width=3.1in]{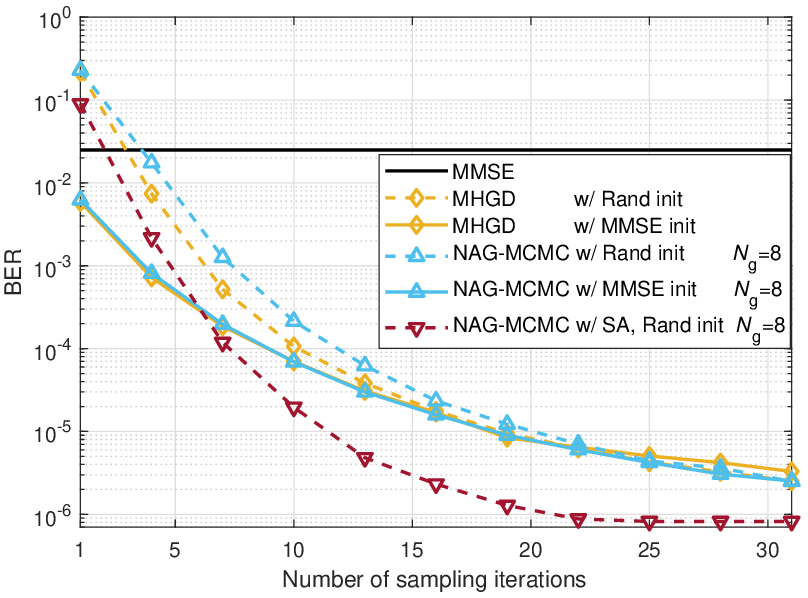}
      \caption{\Times{8}{8} MIMO, 16-QAM, \SNR{=}{25}, $P=16$}
      \label{fig:convergence_8mimo}
    \end{subfigure}
    \begin{subfigure}{3.1in}
      \includegraphics[width=3.1in]{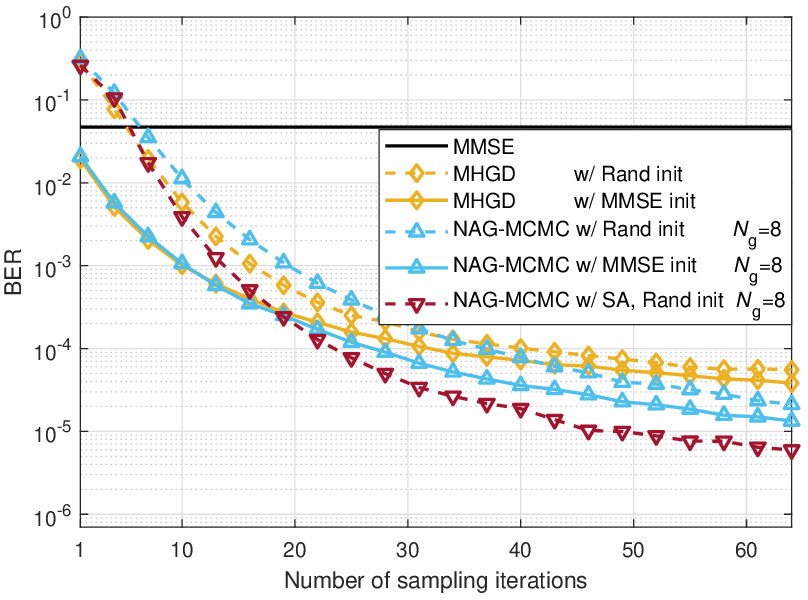}
      \caption{\Times{16}{16} MIMO, 64-QAM, \SNR{=}{30}, $P=32$}
      \label{fig:convergence_16mimo}
    \end{subfigure}
    \vspace{-0.1cm}
    \caption{BER convergence property with respect to the number of sampling {iterations} in 
    Rayleigh fading channels.}
    \label{fig:convergence}
  \end{figure}
}

\CheckRmv{
  \begin{figure}[t]
    \captionsetup[subfigure]{aboveskip=-0.5pt,belowskip=-0.5pt}
    \centering
    \begin{subfigure}{3.1in}
      \includegraphics[width=3.1in]{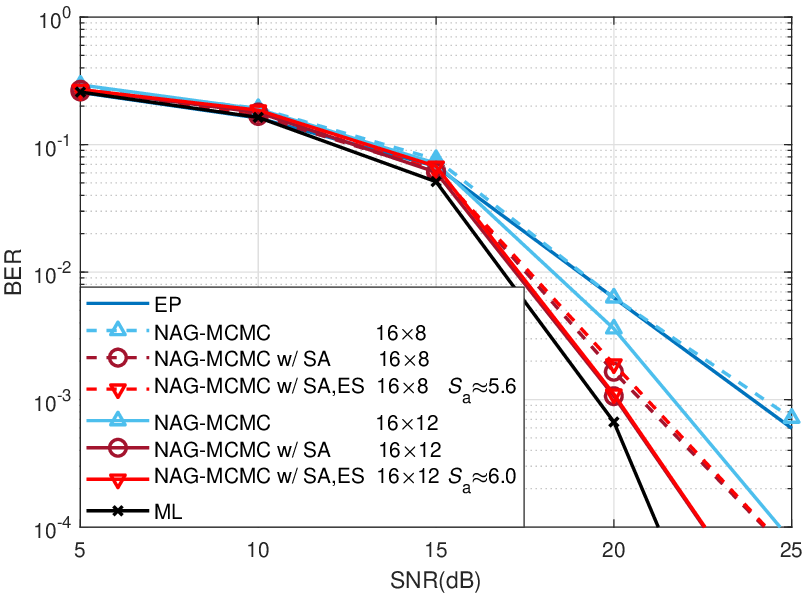}
      \caption{16-QAM}
      \label{fig:ber_8mimo_16qam_1}
    \end{subfigure}
    \begin{subfigure}{3.1in}
      \includegraphics[width=3.1in]{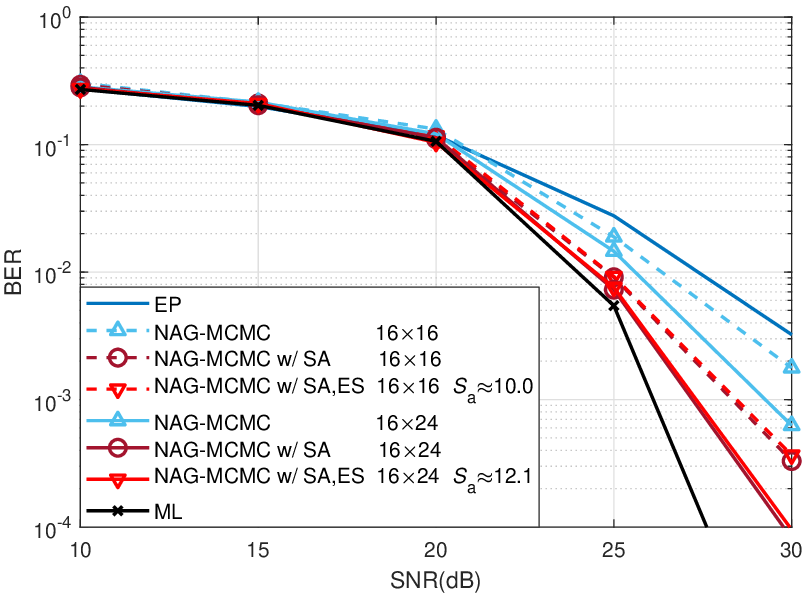}
      \caption{64-QAM}
      \label{fig:ber_8mimo_64qam_1}
    \end{subfigure}
    \vspace{-0.1cm}
    \caption{BER performance comparsion between NAG-MCMC and its enhanced versions in an \Times{8}{8} MIMO system with 16-QAM/64-QAM modulation and Rayleigh fading channels.
    } 
    
    \label{fig:ber_8mimo_1}
  \end{figure}
}
\subsection{Uncoded MIMO Detectors}
In this subsection, we evaluate the uncoded BER performance. 
First, the BER convergence property of the proposed schemes is analyzed.
Then, the performance under medium-sized MIMO channels is given in the form of BER as a function of SNR.
Finally, the effectiveness of the proposed schemes under high-dimensional MIMO systems, realistic channel models, {and channel estimation errors} is verified. 

\subsubsection{Convergence Property}

\CheckRmv{
  \begin{figure*}[t]
    \captionsetup[subfigure]{aboveskip=-0.5pt,belowskip=-0.5pt}
    \centering
    \begin{subfigure}{3.1in}
      \includegraphics[width=3.1in]{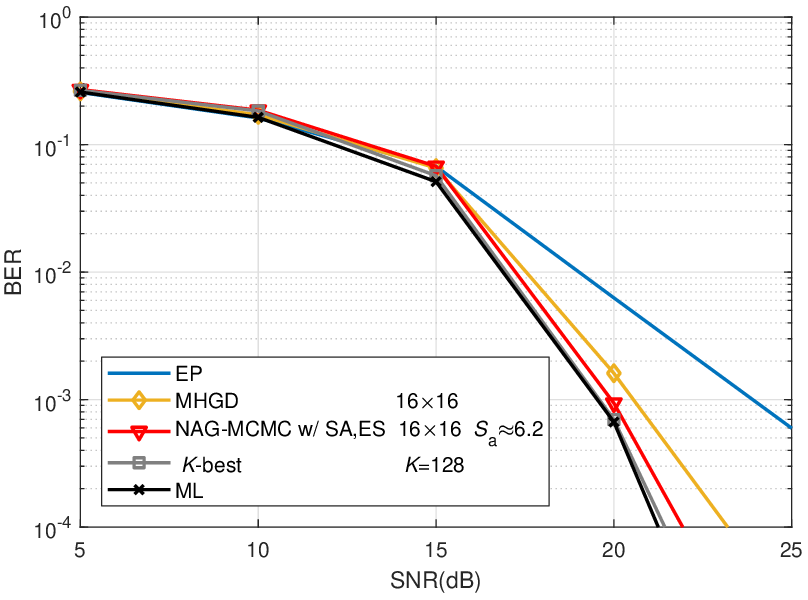}
      \caption{BER performance (16-QAM)}
      \label{fig:ber_8mimo_16qam_2}
    \end{subfigure}
    \begin{subfigure}{3.1in}
      \includegraphics[width=3.1in]{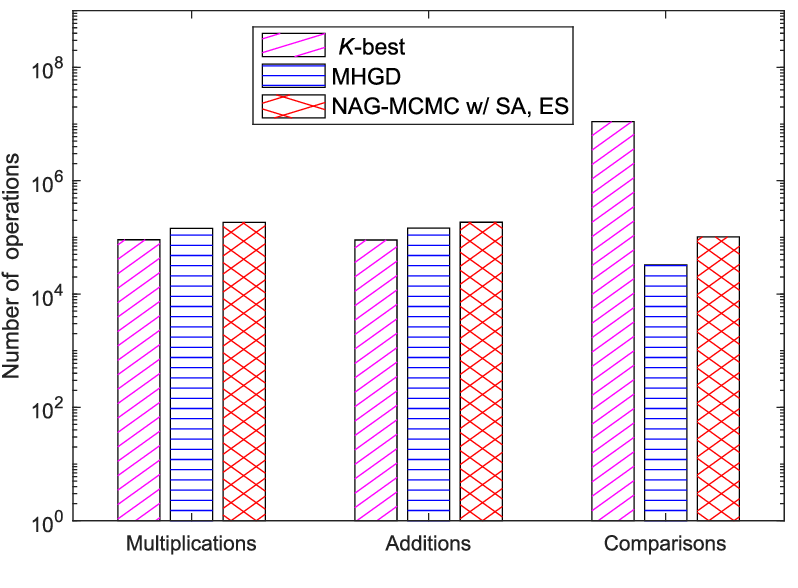}
      \caption{Computational complexity (16-QAM)}
      \label{fig:complexity_8mimo_16qam}
    \end{subfigure}
    \begin{subfigure}{3.1in}
      \includegraphics[width=3.1in]{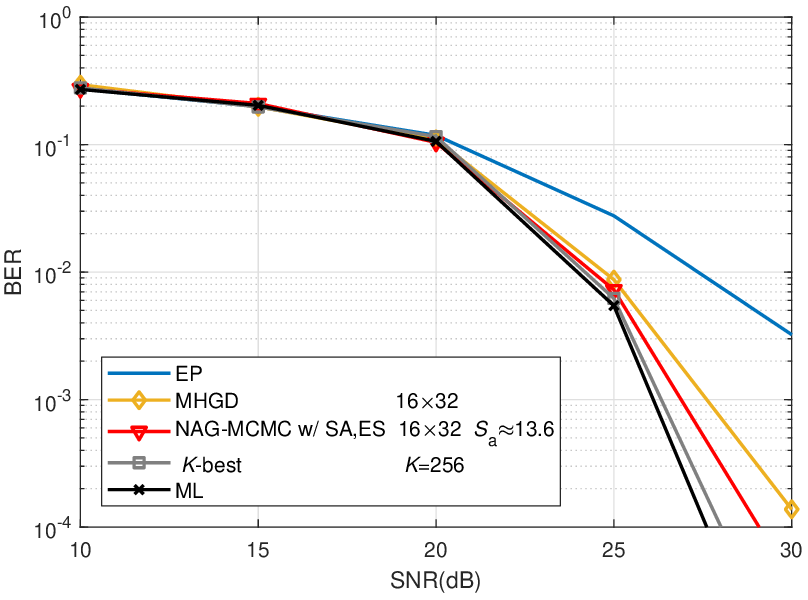}
      \caption{BER performance (64-QAM)}
      \label{fig:ber_8mimo_64qam_2}
    \end{subfigure}
    \begin{subfigure}{3.1in}
      \includegraphics[width=3.1in]{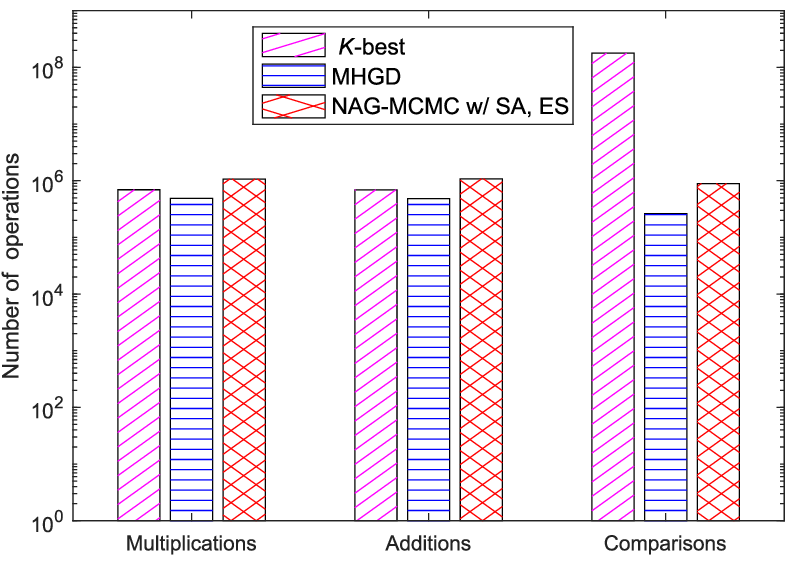}
      \caption{Computational complexity (64-QAM)}
      \label{fig:complexity_8mimo_64qam}
    \end{subfigure}
    \vspace{-0.1cm}
    \caption{{BER performance and computational complexity for an \Times{8}{8} MIMO system with 16-QAM/64-QAM modulation and Rayleigh fading channels.}} 
    \label{fig:ber_8mimo_2}
  \end{figure*}
}
\figref{fig:convergence} shows the convergence property in terms of the BER with respect to the number of sampling iterations, where the performance of the MMSE method is set as a reference. 
Figs.~\ref{fig:convergence}(a) and \ref{fig:convergence}(b) show the results under different dimensions.
In \figref{fig:convergence}(a), where an \Times{8}{8}  
MIMO system with 16-QAM {and \SNR{=}{25}} is considered, the proposed NAG-MCMC with $N_{\rm g}=8$ successive Nesterov's GDs can approach the convergence performance of the MHGD, 
{regardless of the initialization (MMSE or random).} 
For the \Times{16}{16} MIMO system with 64-QAM considered in \figref{fig:convergence}(b), the NAG-MCMC with $N_{\rm g}=8$ even obtains a performance gain over MHGD.
Therefore, the number of successive Nesterov's GDs required by NAG-MCMC to reach the equivalent performance as MHGD does not increase with the MIMO dimensions, which is a notable strength of the proposed method. 
We hence set $N_{\rm g}=8$ for NAG-MCMC and the enhanced versions in the remainder of the simulations.

In another aspect, the comparison between the MMSE and random initialization shows that the MMSE initialization provides  superior performance at the early stage,  
whereas the final performance of the algorithms near convergence under the two initialization is comparable. 
{Furthermore, the NAG-MCMC with SA and 
random initialization significantly outperforms the MHGD and NAG-MCMC, 
reaching the same BER with substantially fewer iterations and only a modest increase of computations in each iteration.} 

We can also conclude from \figref{fig:convergence} that the NAG-MCMC-based algorithms can efficiently navigate the search space. 
Even with the search space dimensions dramatically increasing from $16^8$ in \figref{fig:convergence}(a) to $64^{16}$ in \figref{fig:convergence}(b), the required number of iterations to approach convergence only slightly increases from 30 to 60. 
Moreover, the proposed methods can achieve a BER similar to that of the MMSE with a rather small number of iterations, 
demonstrating their flexibility in adapting to various computational demands.

\subsubsection{Performance under Medium-Sized MIMO Channels}

We investigate the BER performance under a medium-sized MIMO system with $\nr\times \nt=8\times 8$ and different types of modulation,  
using the Rayleigh fading model as the wireless channel.  
{\figref{fig:ber_8mimo_1} illustrates the BER performance comparison between the proposed NAG-MCMC and its enhanced versions.  
The conventional EP and optimal ML are used as baselines.} 
The number of samplers and sampling iterations is denoted by $P\times S$ in the legend. 

In \figref{fig:ber_8mimo_1}(a), the basic NAG-MCMC algorithm with $P=16$ parallel samplers and $S=8$ achieves comparable performance to the EP baseline under 16-QAM.
{Meanwhile, NAG-MCMC with SA and the same  
number of samplers and iterations remarkably improves the BER performance.}  
Specifically, the gain over NAG-MCMC is approximately 4 dB if we target at $\text{BER} =10^{-3}$. 
{The NAG-MCMC with both SA and ES further reduces the number of iterations to $S_{\rm a}\approx5.6$ on average under different SNRs with virtually no performance loss compared to NAG-MCMC with SA but no ES adopted, saving computational resources and promoting implementation efficiency.}
All of the proposed algorithms have a performance gain of approximately 1 dB with the preset number of iterations increasing to $S=12$, and the NAG-MCMC with SA approaches the optimal ML performance in this situation.
{The ES strategy decreases the average number of iterations to $S_{\rm a}\approx6.0$, only half of the preset number.}
The results in \figref{fig:ber_8mimo_1}(b) show a similar trend as those in \figref{fig:ber_8mimo_1}(a) and further verify the effectiveness of the NAG-MCMC and the enhancing strategies under the high-order 64-QAM modulation.

{\figref{fig:ber_8mimo_2} compares the performance and complexity of the enhanced NAG-MCMC to the MHGD baseline  
under the same system settings as \figref{fig:ber_8mimo_1}. 
Both methods  
are configured with the same number of parallel samplers and iterations. 
Additionally, the $K$-best detector with a large list size (denoted by the parameter $K$) serves as a  near-ML baseline and a complexity reference \cite{guoAlgorithmImplementationKbest2006}.
In \figref{fig:ber_8mimo_2}(a), the BER curves under 16-QAM show that both MHGD and the enhanced NAG-MCMC outperform EP by a large margin and approach the performance of $K$-best and ML.
The enhanced NAG-MCMC has 
a performance gain of approximately 1 dB over MHGD at the BER of $10^{-4}$, with the average number of iterations less than half of the preset number. 

To provide a comprehensive complexity comparison in this scenario, the number of multiplication, addition, and comparison operations are shown in \figref{fig:ber_8mimo_2}(b) because the complexity of $K$-best SD is dominated by list-sorting \cite{hochwaldAchievingNearcapacityMultipleantenna2003,guoAlgorithmImplementationKbest2006}, which involves a large number of comparison operations. 
The figure verifies the superiority of the enhanced NAG-MCMC over MHGD in
achieving noticeable performance gains with similar complexity.
Moreover, the proposed method requires significantly fewer comparison operations, approximately two orders of magnitude lower than the $K$-best detector, while the number of other operations remains similar.
Considering that our method further benefits from parallel implementation to achieve low-latency processing, while the $K$-best algorithm is executed sequentially, the enhanced NAG-MCMC demonstrates high promise. 

When the constellation enlarges to 64-QAM, 
the compared methods are equipped with more iterations and a larger list size to enhance performance.
\figref{fig:ber_8mimo_2}(c) shows that NAG-MCMC with SA and ES remarkably outperforms the MHGD and is much closer to ML in this high-order modulation setting.
Furthermore, \figref{fig:ber_8mimo_2}(d) demonstrates that its complexity is substantially lower as compared to $K$-best and comparable to that of MHGD when achieving this performance.}

\subsubsection{Performance with Large-Scale MIMO Systems}

In this subsection, we examine the performance of the proposed NAG-MCMC with SA and ES under MIMO systems with a large number of antennas. 
Specifically, we present the BER performance under \Times{64}{64} and \Times{128}{128} MIMO systems with 16-QAM and Rayleigh fading channels in \figref{fig:ber_64_128mimo}.
{Due to poor performance even with a large $K$ (e.g., $K = 2048$) and prohibitively increasing complexity, we do not provide the $K$-best baseline.}
In \figref{fig:ber_64_128mimo}(a), where a \Times{64}{64} MIMO configuration is considered, the NAG-MCMC with SA and ES retains a substantial performance improvement over EP and MHGD.
Note that the number of parallel samplers in NAG-MCMC with SA and ES is only half of that in MHGD when obtaining this gain.
\figref{fig:ber_64_128mimo}(b) shows that the advantage of NAG-MCMC with SA and ES is still valid in the \Times{128}{128} MIMO system. 
Moreover, the numbers of multiplications cost by MHGD are $5.52 \times 10^7$ and $4.30 \times 10^8$ under the \Times{64}{64} and \Times{128}{128} MIMO systems, respectively. 
By contrast, the counterparts of NAG-MCMC with SA and ES are $4.20 \times 10^7$ and $2.76\times 10^8$, lower than those of the MHGD, especially for the \Times{128}{128} MIMO system.
This complexity comparison demonstrates that NAG-MCMC with SA and ES is more scalable to large-dimensional systems than MHGD. 

% \CheckRmv{
%   \begin{figure}[t]
%     \centering
%     \subfigure[\Times{64}{64} MIMO]{
%       \includegraphics[width=3.1in]{BER_64MIMO_16QAM.eps}
%       \label{fig:ber_64mimo}
%     }
%     \subfigure[\Times{128}{128} MIMO]{
%       \includegraphics[width=3.1in]{BER_128MIMO_16QAM.eps}
%       \label{fig:ber_128mimo}
%     }
%     \vspace{-0.8cm}
%     \caption{BER performance for the \Times{64}{64} and \Times{128}{128} MIMO cases with 16-QAM modulation under Rayleigh fading channels.
%     % comparison between the proposed NAG-MCMC with SA and ES and MHGD under large-scale systems. Parameters: \Times{64}{64} or \Times{128}{128} MIMO, 16-QAM, Rayleigh fading channels.
%     }
%     \label{fig:ber_64_128mimo} % $S_{\max}=16$ for 64 and 128 MIMO
%   \end{figure}
% }

\CheckRmv{
  \begin{figure}[t]
    \captionsetup[subfigure]{aboveskip=-0.5pt,belowskip=-0.5pt}
    \centering
    \begin{subfigure}{3.1in}
      \includegraphics[width=3.1in]{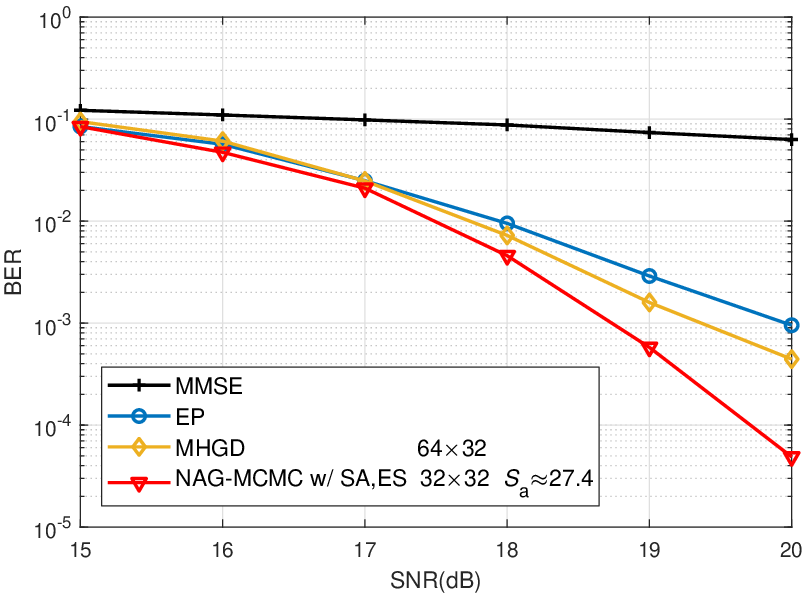}
      \caption{\Times{64}{64} MIMO}
      \label{fig:ber_64mimo}
    \end{subfigure}
    \begin{subfigure}{3.1in}
      \includegraphics[width=3.1in]{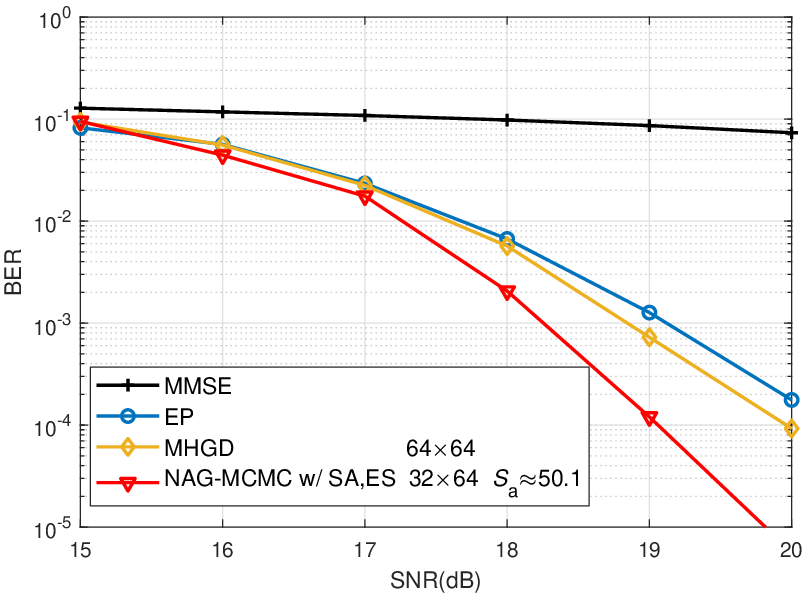}
      \caption{\Times{128}{128} MIMO}
      \label{fig:ber_128mimo}
    \end{subfigure}
    \vspace{-0.15cm}
    \caption{BER performance for large-scale MIMO systems with 16-QAM modulation and Rayleigh fading channels.
    } 
    \label{fig:ber_64_128mimo}
  \end{figure}
}

\subsubsection{Performance under 3GPP MIMO Channels}

To further verify the effectiveness of the proposed method, we evaluate the performance under the 3GPP MIMO channel datasets generated by the QuaDRiGa toolbox \cite{jaeckel2014quadriga}. 
Specifically, we consider the 3GPP urban macrocell non-line-of-sight uplink transmission, where a base station (BS) with 32 dual-polarized antennas (i.e., $\nr=64$) serves 8 single-antenna users uniformly placed within a $120^{\circ}$ and radius-500 m cell sector, giving an effective \Times{64}{8} MIMO system.
The antenna array configuration at the BS is a planar array with half wavelength of antenna spacing installed at the height of 25 m.
The channel is at the center frequency of 2.53 GHz, with 20 MHz bandwidth and 256 effective subcarriers. 
We generate 1200 independent channel realizations, i.e., \Times{1200}{256} channel matrices, with random user locations for performance evaluation. 
The modulation type is 16-QAM.
\CheckRmv{
  \begin{figure}[t]
    \setlength{\abovecaptionskip}{-0.1cm}
    \centering
    \includegraphics[width=3.1in]{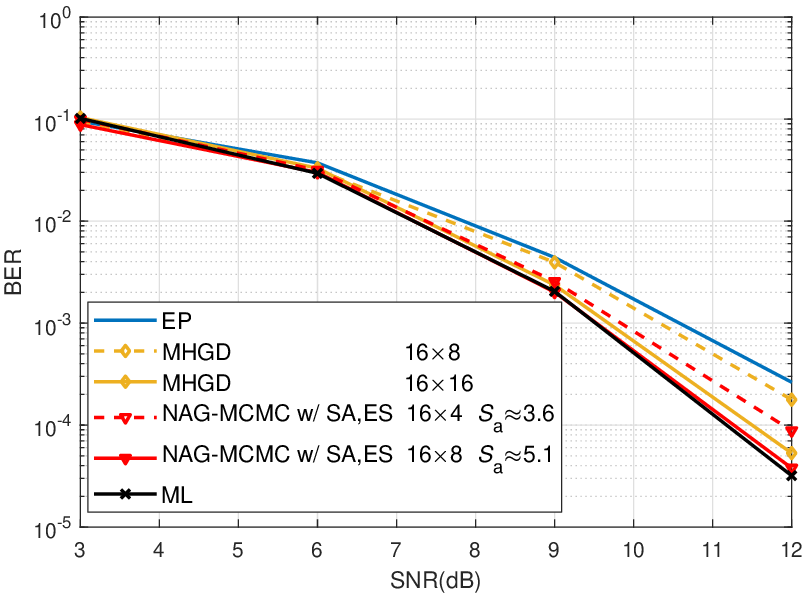}
    \caption{BER performance comparison between MHGD and NAG-MCMC with SA and ES under the 3GPP channel model with 16-QAM modulation in a \Times{64}{8} MIMO system.
    }
    \label{fig:ber_3gpp}  
  \end{figure}
}

\figref{fig:ber_3gpp} shows the BER performance under the 3GPP channel model.
The number of samplers is set as $P=16$ for MHGD and NAG-MCMC with SA and ES in this scenario.
{The NAG-MCMC with SA and ES under very few sampling iterations ($S=4$ and $S_{\rm a}=3.6$) still outperforms EP and the MHGD with $S=8$ by a large margin.} 
Moreover, the NAG-MCMC with SA and ES can further approach the ML performance with $S$ increasing to 8, where the gap is less than 0.2 dB, and the average number of sampling iterations is still maintained at a low level of $S_{\rm a}\approx 5.1$.
The NAG-MCMC with SA and ES under this parameter setting is also superior to the MHGD with $S=16$, revealing its high efficiency and robustness towards realistic channels.  

\CheckRmv{
  \begin{figure}[t]
    \setlength{\abovecaptionskip}{-0.1cm}
    \centering
    \includegraphics[width=3.1in]{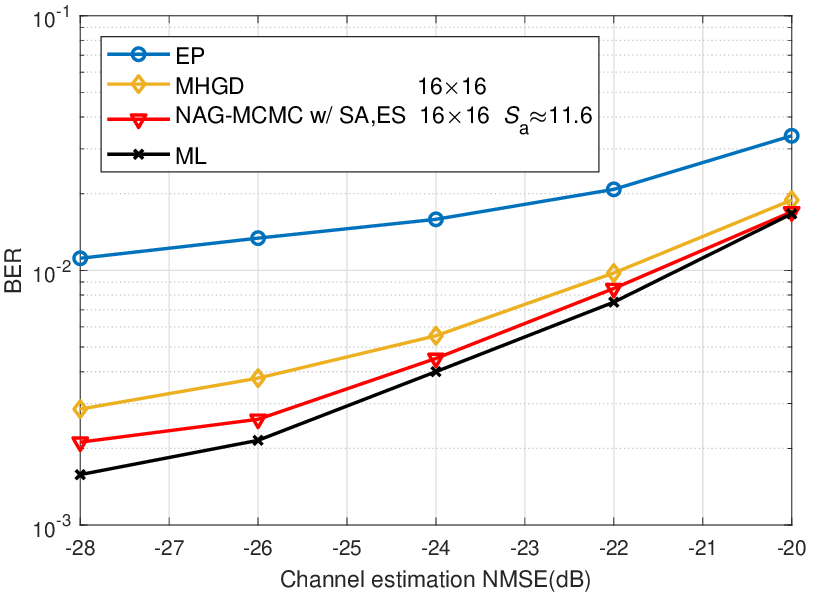}
    \caption{{BER performance  
    as a function of channel estimation NMSE
    in an \Times{8}{8} MIMO system with \SNR{=}{20}, 16-QAM modulation, and Rayleigh fading channels.}}
    \label{fig:ber_8mimo_ce}
  \end{figure}
}
\subsubsection{{Performance under Channel Estimation Errors}}
{In our previous investigations, perfect CSI at the receiver was assumed. 
Now, we evaluate the proposed detector's performance with channel estimation errors. 
We model the errors using a linear MMSE channel estimator \cite{FundamentalsStatisticalSignal}. 
The estimated channel matrix is given by:
\CheckRmv{
  \begin{equation}
    \hat{\mathbf{H}} = \mathbf{H} + \Delta \mathbf{H},
  \end{equation}
}
where  
$\mathbf{H}$ is the true channel matrix, and $\Delta \mathbf{H}$ is the estimation error matrix.
Each element of  $\Delta \mathbf{H}$ is independent and identically Gaussian-distributed, i.e., $(\Delta \mathbf{H})_{ij}\sim\mathcal{CN}(0, \sigma_h^2)$, where $\sigma_h^2$ is the variance of the channel estimation errors \cite{weberImperfectChannelstateInformation2006}.
We define the channel estimation normalized mean square error (NMSE) as 
\CheckRmv{
  \begin{equation}
    \text{NMSE} = \frac{\mathbb{E}[\|\Delta \mathbf{H}\|_F^2]}{\mathbb{E}[\| \mathbf{H}\|_F^2]}
  \end{equation}
} 
and evaluate the BER performance of different detectors as a function of this metric.
In an \Times{8}{8} MIMO system using 16-QAM modulation, operating under Rayleigh fading channels with an SNR of 20 dB, 
\figref{fig:ber_8mimo_ce} shows the BER performance of various detectors concerning channel estimation errors. 
The proposed NAG-MCMC with SA and ES consistently outperforms the EP and MHGD detectors and approaches the performance of ML under different levels of channel estimation accuracy, demonstrating its robustness against channel estimation errors.}

\subsection{Coded MIMO Receivers}

\CheckRmv{
  \begin{figure}[t]
    \captionsetup[subfigure]{aboveskip=-0.5pt,belowskip=-0.5pt}
    \centering
    \begin{subfigure}{3.1in}
      \includegraphics[width=3.1in]{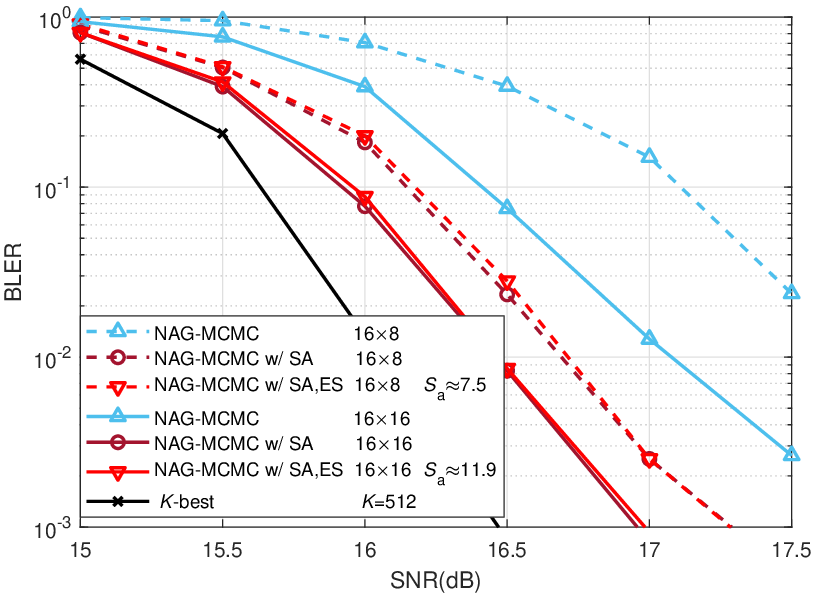}
      \caption{16-QAM}
      \label{fig:per_8mimo_16qam_ldpc_1}
    \end{subfigure}
    \begin{subfigure}{3.1in}
      \includegraphics[width=3.1in]{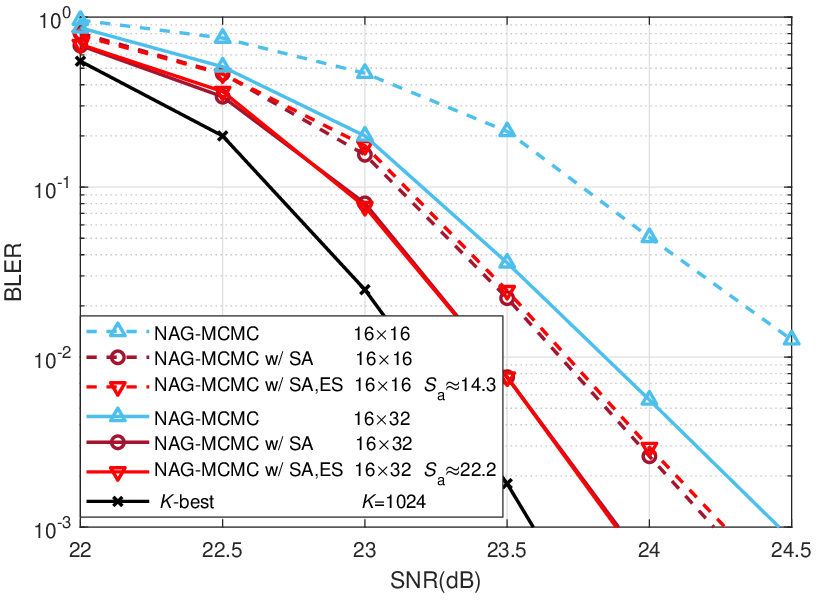}
      \caption{64-QAM}
      \label{fig:per_8mimo_64qam_ldpc_1}
    \end{subfigure}
    \vspace{-0.1cm}
    \caption{BLER performance comparison between NAG-MCMC and its enhanced versions in an \Times{8}{8} MIMO system with LDPC codes (rate-3/4, block length-1944), 16-QAM/64-QAM modulation, and Rayleigh fading channels.} 
    \label{fig:per_8mimo_ldpc_1}
  \end{figure}
}

In this subsection, we examine an \Times{8}{8} LDPC-coded MIMO system. 
The specifications of the LDPC codes are given in \secref{sec:parameter}.
The simulated channels follow the Rayleigh fading model.
\figref{fig:per_8mimo_ldpc_1} presents the BLER performance of the proposed methods under 16-QAM and 64-QAM. 
In \figref{fig:per_8mimo_ldpc_1}(a) where 16-QAM is considered,  
the BLER curves demonstrate that the NAG-MCMC with SA significantly improves the performance of the NAG-MCMC in coded MIMO systems. 
{Meanwhile, the ES strategy is still valid in the coded MIMO detector.
Specifically, for the $P\times S =16\times 16$ configuration, the performance of NAG-MCMC with both SA and ES versus NAG-MCMC with SA only shows that the former reduces the average number of iterations to $S_{\rm a}\approx 11.9$ and incurs no perceivable performance loss.
Furthermore, the performance gap between the NAG-MCMC with  SA and ES under $P\times S =16\times 16$ and the near-ML but computationally expensive $K$-best detector \cite{guoAlgorithmImplementationKbest2006} is only about 0.5 dB.}
Moreover, \figref{fig:per_8mimo_ldpc_1}(b) shows similar results 
as those in \figref{fig:per_8mimo_ldpc_1}(a),  
demonstrating that our proposed methods also perform well in coded systems with high-order modulation.

% \CheckRmv{
% \begin{figure}[t]
%   \centering
%   \subfigure[16-QAM]{
%       \includegraphics[width=3.1in]{CODED_FER_8MIMO_16QAM_LDPC_2.eps}
%       \label{fig:per_8mimo_16qam_ldpc_2}
%     }
%     \subfigure[64-QAM]{
%       \includegraphics[width=3.1in]{CODED_FER_8MIMO_64QAM_LDPC_2.eps}
%       \label{fig:per_8mimo_64qam_ldpc_2}
%     }
%   \vspace{-0.8cm}
%   \caption{BLER performance comparison between MHGD and NAG-MCMC with SA and ES in an \Times{8}{8} MIMO system with LDPC codes (rate-3/4, block length-1944), 16-QAM/64-QAM modulation, and Rayleigh fading channels.} % No Smax is set
%   \label{fig:per_8mimo_ldpc_2}
% \end{figure}
% }

\CheckRmv{
  \begin{figure}[t]
    \captionsetup[subfigure]{aboveskip=-0.5pt,belowskip=-0.5pt}
    \centering
    \begin{subfigure}{3.1in}
      \includegraphics[width=3.1in]{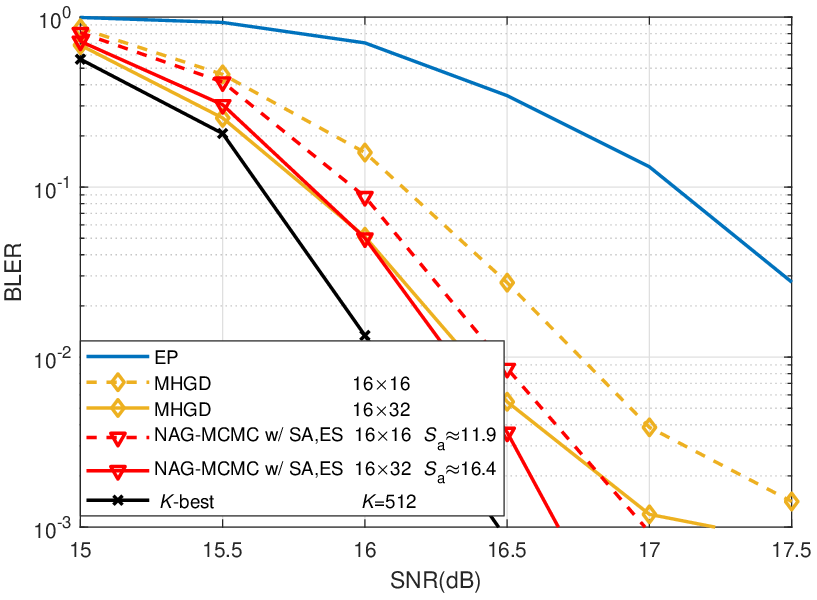}
      \caption{16-QAM}
      \label{fig:per_8mimo_16qam_ldpc_2}
    \end{subfigure}
    \begin{subfigure}{3.1in}
      \includegraphics[width=3.1in]{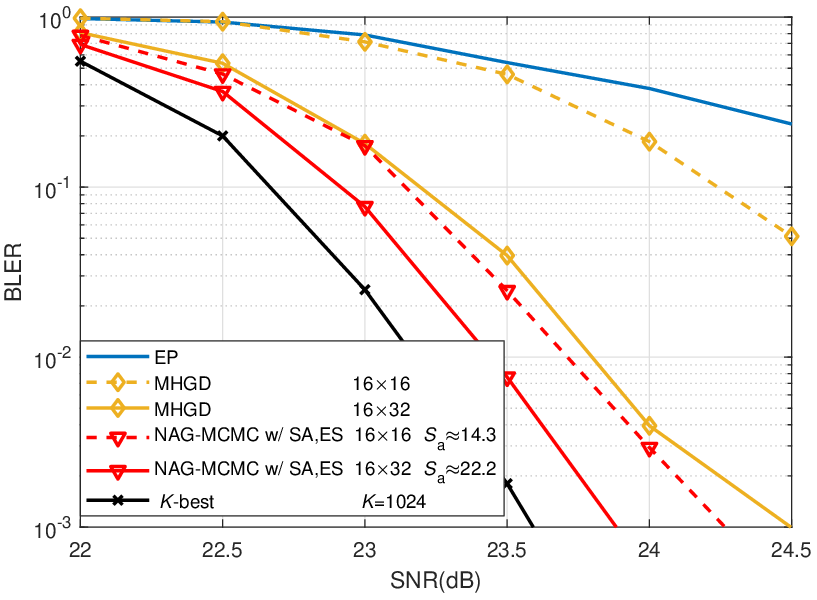}
      \caption{64-QAM}
      \label{fig:per_8mimo_64qam_ldpc_2}
    \end{subfigure}
    \vspace{-0.1cm}
    \caption{BLER performance comparison between MHGD and NAG-MCMC with SA and ES in an \Times{8}{8} MIMO system with LDPC codes (rate-3/4, block length-1944), 16-QAM/64-QAM modulation, and Rayleigh fading channels.} 
    \label{fig:per_8mimo_ldpc_2}
  \end{figure}
}

\figref{fig:per_8mimo_ldpc_2} provides a comparison of the coded BLER performance of the MHGD and our proposed NAG-MCMC with SA and ES under the same system settings as \figref{fig:per_8mimo_ldpc_1}. 
{We can find from \figref{fig:per_8mimo_ldpc_2}(a) that under the settings of both $P\times S=16\times 16$ and $P\times S=16\times 32$, the NAG-MCMC with SA and ES outperforms the MHGD.}
Furthermore, the NAG-MCMC with SA and ES under $P\times S =16\times 32$ narrows the gap with the $K$-best to merely 0.2 dB at $\text{BLER}=10^{-3}$.
Meanwhile, the ES strategy reduces the number of iterations from the preset $S=32$ to the substantially smaller $S_{\rm a}\approx 16.4$ on average. 
{For complexity comparison, the number of complex multiplications required by MHGD and NAG-MCMC with SA and ES under $P\times S =16\times 32$ is $2.88\times 10^5$ and $4.78\times 10^5$, respectively.}
{Finally, the results in \figref{fig:per_8mimo_ldpc_2}(b) further show that the improvement of the NAG-MCMC with SA and ES over the MHGD is impressive under the higher-order modulation, where the NAG-MCMC with SA and ES under $P\times S =16\times 16$ even outperforms the MHGD under $P\times S=16\times 32$ at all evaluated SNRs.}
In addition, when the preset number of iterations increases to $S=32$, the gain of NAG-MCMC with SA and ES gets much more pronounced.
{The number of complex multiplications in this case is $1.74\times 10^6$ for NAG-MCMC with SA and ES, as compared with $4.87\times 10^5$ for MHGD.}

%%%%%%%%%%%%%%%%%%%%%%%%%%%%%%%%%%%%%%%%%%%%%%%%%%%%%%%%%%
\section{Conclusion}
We proposed a gradient-based MCMC method, named NAG-MCMC, for MIMO detection.
NAG-MCMC seamlessly combines the strengths of the MCMC and Nesterov's GD to create a {high-performance} and efficient exploration of the search space for MIMO detection. 
By operating in a \textit{multiple GDs per random walk} style, NAG-MCMC avoids the burdensome Newton's method while still delivering high performance. 
On this basis, we designed the SA strategy to achieve a finer-grained exploration of the search space for performance improvement with a very small complexity increase. 
Furthermore, our proposed ES strategy substantially reduces unnecessary searches and thus leads to a noticeable complexity reduction. 
Simulation results demonstrate the superiority of the proposed detectors over the conventional and state-of-the-art MIMO detectors, achieving near-ML performance in both uncoded and coded MIMO systems.
Complexity analysis further demonstrates the prominent strength of the proposed schemes in scalability to high dimensions, revealing its potential for use in large-scale MIMO systems.  
%%%%%%%%%%%%%%%%%%%%%%%%%%%%%%%%%%%%%%%%%%%%%%%%%%%%%%%%%%
% \appendices
\appendix[{Proof of Lemma~\ref{lemma2}}]  \label{sec:appendix}
% \vspace{-0.3cm}
{The property of convexity is naturally satisfied because $f(\mathbf{x})=\frac{1}{2}\|\mathbf{y}-\mathbf{Hx}\|^2$ (in the continuous complex space) is a quadratic function \cite{boyd2004convex}.
    We next prove the $L$-Lipschitz property of the gradient $\nabla f(\mathbf{x})$.
		For any $\mathbf{x}_1, \mathbf{x}_2 \in \mathbb{C}^{\nt \times 1}$, we have
		\CheckRmv{
		  \begin{align}
		   \|\nabla f(\mathbf{x}_1) -&\nabla f(\mathbf{x}_2)\|^2 \nonumber\\ 
        =& \| \mathbf{H}^{\rm H}\mathbf{H}(\mathbf{x}_1-\mathbf{x}_2)\|^2 \nonumber\\
		    =&\big(\mathbf{U}\boldsymbol{\Sigma}\mathbf{U}^{\rm H}(\mathbf{x}_1-\mathbf{x}_2)\big)^{\rm H}\big(\mathbf{U}\boldsymbol{\Sigma}\mathbf{U}^{\rm H}(\mathbf{x}_1-\mathbf{x}_2)\big),
		  \end{align}
		}
		where $\mathbf{U}\boldsymbol{\Sigma}\mathbf{U}^{\rm H}$ is the eigenvalue decomposition of $\mathbf{H}^{\rm H}\mathbf{H}$, and $\boldsymbol{\Sigma}$ is a diagonal matrix whose diagonal elements are the eigenvalues $\{\lambda_1,\ldots,\lambda_{\nt}\}$ of $\mathbf{H}^{\rm H}\mathbf{H}$ arranged in  descending order.
		Since $\mathbf{U}^{\rm H}\mathbf{U}=\mathbf{I}_{\nt}$, we have
		\CheckRmv{
		  \begin{equation}
		    \|\nabla f(\mathbf{x}_1)-\nabla f(\mathbf{x}_2)\|^2 =(\mathbf{x}_1-\mathbf{x}_2)^{\rm H}\mathbf{U}\boldsymbol{\Sigma}^2
		    \mathbf{U}^{\rm H}(\mathbf{x}_1-\mathbf{x}_2), 
		    \label{eq:evd}
		  \end{equation}
		}
		Denote the maximum eigenvalue of $\mathbf{U}\boldsymbol{\Sigma}^2 \mathbf{U}^{\rm H}$ as $\bar{\lambda}_{\max}$. 
		Using the upper bound of Rayleigh quotient, we have
		\CheckRmv{
		  \begin{equation}
		    \frac{(\mathbf{x}_1-\mathbf{x}_2)^{\rm H}\mathbf{U}\boldsymbol{\Sigma}^2
		    \mathbf{U}^{\rm H}(\mathbf{x}_1-\mathbf{x}_2)}{(\mathbf{x}_1-\mathbf{x}_2)^{\rm H}(\mathbf{x}_1-\mathbf{x}_2)}\leq \bar{\lambda}_{\max} = {\lambda}_{\max}^2,
		    \label{eq:rayleigh_quotient}
		  \end{equation}
		}
		where $\lambda_{\max} = \lambda_1 \geq 0$ is the maximum eigenvalue of $\mathbf{H}^{\rm H}\mathbf{H}$.
		From \eqref{eq:evd} and \eqref{eq:rayleigh_quotient}, we have
		\CheckRmv{
		  \begin{align}
		    \|\nabla f(\mathbf{x}_1)-\nabla f(\mathbf{x}_2)\|&\leq \sqrt{\lambda_{\max}^2\|\mathbf{x}_1-\mathbf{x}_2\|^2} \nonumber\\
        &= \lambda_{\max}\|\mathbf{x}_1-\mathbf{x}_2\|.
		  \end{align}
		}
		Hence, $\nabla f{(\mathbf{x})}$ is Lipschitz continuous and the Lipschitz constant is $L=\lambda_{\max}$, which is the maximum eigenvalue of $\mathbf{H}^{\rm H}\mathbf{H}$. 
    This completes the proof.}

% \section*{Acknowledgment}


\ifCLASSOPTIONcaptionsoff
  \newpage
\fi


% \bibliographystyle{IEEEtran}      
% \bibliography{IEEEabrv,ref}

\begin{thebibliography}{10}
  \providecommand{\url}[1]{#1}
  \csname url@samestyle\endcsname
  \providecommand{\newblock}{\relax}
  \providecommand{\bibinfo}[2]{#2}
  \providecommand{\BIBentrySTDinterwordspacing}{\spaceskip=0pt\relax}
  \providecommand{\BIBentryALTinterwordstretchfactor}{4}
  \providecommand{\BIBentryALTinterwordspacing}{\spaceskip=\fontdimen2\font plus
  \BIBentryALTinterwordstretchfactor\fontdimen3\font minus
    \fontdimen4\font\relax}
  \providecommand{\BIBforeignlanguage}[2]{{%
  \expandafter\ifx\csname l@#1\endcsname\relax
  \typeout{** WARNING: IEEEtran.bst: No hyphenation pattern has been}%
  \typeout{** loaded for the language `#1'. Using the pattern for}%
  \typeout{** the default language instead.}%
  \else
  \language=\csname l@#1\endcsname
  \fi
  #2}}
  \providecommand{\BIBdecl}{\relax}
  \BIBdecl
  
  \bibitem{bjornson2017massive}
  E.~Bj{\"o}rnson, J.~Hoydis, and L.~Sanguinetti, ``Massive MIMO
    networks: Spectral, energy, and hardware efficiency,'' \emph{Foundations and
    Trends{\textregistered} in Signal Processing}, vol.~11, no. 3-4, pp.
    154--655, 2017.

  \bibitem{ngoEnergySpectralEfficiency2013}
  H.~Q. Ngo, E.~G. Larsson, and T.~L. Marzetta, ``Energy and {{spectral
    efficiency}} of {{very large multiuser MIMO systems}},'' \emph{IEEE Trans. Commun.}, vol.~61, no.~4, pp. 1436--1449, Apr. 2013.
  
  \bibitem{yangFiftyYearsMIMO2015}
  S.~Yang and L.~Hanzo, ``Fifty {{years}} of {{MIMO detection}}: {{The road}} to
    {{large-scale MIMOs}},'' \emph{IEEE Commun. Surv. Tuts.}, vol.~17, no.~4,
    pp. 1941--1988, 2015.
  
  \bibitem{albreemMassiveMIMODetection2019}
  M.~A. Albreem, M.~Juntti, and S.~Shahabuddin, ``Massive {{MIMO detection
    techniques}}: {{A survey}},'' \emph{IEEE Commun. Surv. Tuts.}, vol.~21,
    no.~4, pp. 3109--3132, 2019.
  
  \bibitem{hochwaldAchievingNearcapacityMultipleantenna2003}
  B.~Hochwald and S.~{ten Brink}, ``Achieving near-capacity on a multiple-antenna
    channel,'' \emph{IEEE Trans. Commun.}, vol.~51, no.~3, pp.
    389--399, Mar. 2003.
  
  \bibitem{guoAlgorithmImplementationKbest2006}
  Z.~Guo and P.~Nilsson, ``Algorithm and implementation of the {{K-best}} sphere
    decoding for {{MIMO}} detection,'' \emph{IEEE J. Sel. Areas
    Commun.}, vol.~24, no.~3, pp. 491--503, Mar. 2006.
  
  \bibitem{eldar2022machine}
  Y.~C. Eldar, A.~Goldsmith, D.~G{\"u}nd{\"u}z, and H.~V. Poor, \emph{Machine
    learning and wireless communications}.\hskip 1em plus 0.5em minus 0.4em\relax
    Cambridge University Press, 2022.
  
  \bibitem{bishopPatternRecognitionMachine}
  C.~M. Bishop, \emph{Pattern {{recognition}} and {{machine learning}}}.\hskip 1em plus 0.5em minus 0.4em\relax New York,
  NY, USA: Springer, 2006.
  
  \bibitem{donohoMessagepassingAlgorithmsCompressed2009}
D.~L. Donoho, A.~Maleki, and A.~Montanari,
		``{Message-passing algorithms for compressed
		sensing},'' \emph{{Proc. Nat. Acad. Sci.}}, vol. 106, no.~45, pp.
		18\,914--18\,919, Nov. 2009.
  
  \bibitem{cespedesExpectationPropagationDetection2014}
  J.~C{\'e}spedes, P.~M. Olmos, M.~{S{\'a}nchez-Fern{\'a}ndez}, and
    F.~{Perez-Cruz}, ``Expectation {{propagation detection}} for {{high-order
    high-dimensional MIMO systems}},'' \emph{IEEE Trans.
    Commun.}, vol.~62, no.~8, pp. 2840--2849, Aug. 2014.
  
  \bibitem{heModelDrivenDeepLearning2020}
  H.~He, C.-K. Wen, S.~Jin, and G.~Y. Li, ``Model-{{driven deep learning}} for
    {{MIMO detection}},'' \emph{IEEE Trans. Signal Process.}, vol.~68, pp.
    1702--1715, Mar. 2020.
  
  \bibitem{kosasihGraphNeuralNetwork2022}
  A.~Kosasih, V.~Onasis, V.~Miloslavskaya, W.~Hardjawana, V.~Andrean, 
    and B.~Vucetic, ``Graph {{neural network aided}} {{MU-MIMO detectors}},'' \emph{IEEE J. Sel. Areas
    Commun.}, vol.~40, no.~9, pp. 2540--2555, Sep. 2022.

  
  \bibitem{nealMCMCUsingHamiltonian2011}
  S.~Brooks, A.~Gelman, G.~Jones, and X.-L. Meng, \emph{Handbook of {{Markov chain Monte Carlo}}.} 
    \hskip 1em plus 0.5em minus 0.4em\relax Boca Raton, FL, USA: {Chapman \& Hall}, 2011.
  
  \bibitem{farhang-boroujenyMarkovChainMonte2006}
  B.~{Farhang-Boroujeny}, H.~Zhu, and Z.~Shi, ``Markov chain {{Monte Carlo}}
    algorithms for {{CDMA}} and {{MIMO}} communication systems,'' \emph{IEEE
    Trans. Signal Process.}, vol.~54, no.~5, pp.~1896--1909, May 2006.
  
  \bibitem{hedstromAchievingMAPPerformance2017}
  J.~C. Hedstrom, C.~H. Yuen, R.-R. Chen, and B.~{Farhang-Boroujeny}, ``Achieving
    {{near MAP performance with}} an {{excited Markov chain Monte Carlo MIMO
    detector}},'' \emph{IEEE Trans. Wireless Commun.}, vol.~16, no.~12, pp.
    7718--7732, Dec. 2017.
  
  \bibitem{dattaNovelMonteCarloSamplingBasedReceiver2013}
  T.~Datta, N.~A. Kumar, A.~Chockalingam, and B.~S. Rajan, ``A {{novel
    Monte-Carlo-sampling-based receiver}} for {{large-scale uplink multiuser MIMO
    systems}},'' \emph{IEEE Trans. Veh. Technol.}, vol.~62,
    no.~7, pp. 3019--3038, Sep. 2013.
  
  \bibitem{huangMCMCDecodingLDPC2020}
  J.-T. Huang and Y.-H. Kim, ``{{MCMC decoding}} of {{LDPC codes}} with {{BP
    preprocessing}},'' in \emph{{{Proc. IEEE Global
    Commun. Conf. (GLOBECOM)}}}, Dec. 2020, pp. 1--5.
  
  \bibitem{wangOvertheAirAntennaArray2022}
  M.~Wang, J.~Chen, J.~Tao, and H.~Li, ``Over-the-{{air antenna array
    calibration}} for {{mmWave hybrid beamforming systems based}} on {{Monte
    Carlo Markov chain method}},'' \emph{IEEE Trans. Veh.
    Technol.}, vol.~72,
    no.~5, pp. 6068--6079, May 2023.
  
  \bibitem{baiLargeScaleMIMODetection2016}
  L.~Bai, T.~Li, J.~Liu, Q.~Yu, and J.~Choi, ``Large-{{scale MIMO detection using
    MCMC approach with blockwise sampling}},'' \emph{IEEE Trans.
    Commun.}, vol.~64, no.~9, pp. 3697--3707, Sep. 2016.

  \bibitem{larawayImplementationMarkovChain2009}
  S.~A. Laraway and B.~{Farhang-Boroujeny}, ``Implementation of a {{Markov chain
    Monte Carlo based multiuser}}/{{MIMO detector}},'' \emph{IEEE Trans.
    Circuits and Syst. I: Reg. Papers}, vol.~56, no.~1, pp. 246--255, Jan. 2009.
  
  \bibitem{hastings1970monte}
  W.~K. Hastings, ``Monte Carlo sampling methods using Markov chains and their
    applications,'' \emph{Biometrika}, vol. 57, no. 1, pp. 97--109, Apr. 1970.
  
  \bibitem{maSamplingCanBe2019}
  Y.-A. Ma, Y.~Chen, C.~Jin, N.~Flammarion, and M.~I. Jordan, ``Sampling can be
    faster than optimization,'' \emph{{Proc. Nat. Acad. Sci.}}, vol. 116, no.~42, pp. 20\,881--20\,885, Sep. 2019.
  
  \bibitem{wellingBayesianLearningStochastic}
  M.~Welling and Y.~W. Teh, ``Bayesian {{learning}} via {{stochastic gradient
    Langevin dynamics}},'' in \emph{Proc. 28th Int. Conf. Mach. Learn. (ICML)}, pp. 681-688, 2011.
  
  \bibitem{gowdaMetropolisHastingsRandomWalk2021}
  N.~M. Gowda, S.~Krishnamurthy, and A.~Belogolovy, ``Metropolis-{{Hastings
    random walk}} along the {{gradient descent direction}} for {{MIMO
    detection}},'' in \emph{Proc. {{IEEE Int. Conf.}}
    {{Commun.}} ({{ICC}})}, {Montreal, QC, Canada}, Jun. 2021, pp. 1--7.
  
  \bibitem{wuStochasticGradientLangevin2022}
  Z.~Wu and H.~Li, ``Stochastic {{gradient Langevin dynamics}} for {{massive MIMO
    detection}},'' \emph{IEEE Commun. Lett.}, vol.~26, no.~5, pp. 1062--1065, May 2022.
  
  \bibitem{zilbersteinAnnealedLangevinDynamics2022}
  N.~Zilberstein, C.~Dick, R.~{Doost-Mohammady}, A.~Sabharwal, and S.~Segarra,
    ``Annealed {{Langevin dynamics}} for {{massive MIMO detection}},'' \emph{IEEE Trans. Wireless Commun.}, vol.~22, no.~6, pp. 3762--3776, Jun. 2023.
  
  
  \bibitem{boyd2004convex}
  S.~Boyd and L.~Vandenberghe, \emph{Convex optimization}.\hskip 1em
    plus 0.5em minus 0.4em\relax Cambridge University Press, 2004.

  \bibitem{wuMinibatchMetropolisHastingsMCMC2019}
  T.-Y. Wu, Y.~X.~R. Wang, and W.~H. Wong, ``Mini-batch {{Metropolis-Hastings
    MCMC}} with {{reversible SGLD proposal}},'' Aug. 2019. [Online] Available: \url{http://arxiv.org/abs/1908.02910}.

  {\bibitem{wangMarkovChainMonte2023}
  B. Wang \emph{et al.}, ``Markov chain Monte Carlo based MIMO detection with reduced complexity,'' in \emph{Proc. IEEE/CIC Int. Conf. Commun. China (ICCC)}, Aug. 2023, pp. 1--6.}

  {\bibitem{dattaRandomRestartReactiveTabu2010}
  T. Datta, N. Srinidhi, A. Chockalingam, and B. S. Rajan, ``Random restart reactive tabu search algorithm for detection in large-MIMO
    systems,'' \emph{IEEE Commun. Lett.}, vol. 14, no. 12, pp. 1107--1109,
    Dec. 2010.}

  {\bibitem{zhaoTabuSearchDetection2007}
  H. Zhao, H. Long, and W. Wang, ``Tabu search detection for MIMO systems,'' in \emph{Proc. IEEE 18th Int. Symp. Pers., Indoor Mobile Radio Commun. (PIMRC)}, Sep. 2007, pp. 1--5.}

  {\bibitem{nguyenDeepLearningAidedTabu2020}
  N. T. Nguyen and K. Lee, ``Deep learning-aided tabu search detection for
  large MIMO systems,'' \emph{IEEE Trans. Wireless Commun.}, vol. 19, no. 6,
  pp. 4262--4275, Jun. 2020.}
  
  \bibitem{nesterov1983method}
  Y.~Nesterov, ``A method for solving the convex programming problem with
    convergence rate o (1/k\^{} 2),'' in \emph{Dokl. Akad. Nauk SSSR}, vol. 269, pp. 543--547,
    1983.
  
  
  
  \bibitem{polyak1964some}
  B.~T. Polyak, ``Some methods of speeding up the convergence of iteration
    methods,'' \emph{USSR Comput. Math. Math. Phys.}, vol.~4, no.~5, pp. 1--17,
    1964.

  {\bibitem{jinAcceleratedGradientDescent2018}
  C.~Jin, P.~Netrapalli, and M.~I. Jordan, ``Accelerated {{gradient descent
    escapes saddle points faster}} than {{gradient descent}},'' in
    \emph{Proc. 31st {{Conf. Learn. Theory}}}, Jul. 2018, pp. 1042--1085.}
  


\bibitem{nesterovIntroductoryLecturesConvex2004}
  Y.~Nesterov, \emph{Introductory lectures on convex optimization}.\hskip 1em
  plus 0.5em minus 0.4em\relax Springer Science+Business Media, 2004.

  {\bibitem{ruderOverviewGradientDescent2017}
  S. Ruder, ``An overview of gradient descent optimization algorithms,'' Sep. 2016. [Online] Available: \url{http://arxiv.org/abs/1609.04747}.}

  \bibitem{barbuHamiltonianLangevinMonte2020}
  A.~Barbu and S.-C. Zhu, ``Hamiltonian and {{Langevin Monte Carlo}},'' in
    \emph{Monte {{Carlo methods}}}, A.~Barbu and S.-C. Zhu, Eds.\hskip 1em plus
    0.5em minus 0.4em\relax {Singapore}: {Springer}, 2020, pp. 281--325.
  
  
  \bibitem{robertson1995comparison}
  P.~Robertson, E.~Villebrun, and P.~Hoeher, ``A comparison of optimal and
    sub-optimal {MAP} decoding algorithms operating in the log domain,'' in
    \emph{Proc. IEEE Int. Conf. Commun. (ICC)},
    Jun. 1995, pp. 1009--1013.
  
  \bibitem{hedstrom2021capacity}
  J.~C. Hedstrom, A.~Rezazadehreyhani, C.~H. Yuen, and B.~Farhang-Boroujeny, ``A
    capacity achieving {MIMO} detector based on stochastic sampling,'' \emph{IEEE Open J. Commun.
    Society}, vol.~2, pp. 2436--2448, Oct. 2021.
  
  \bibitem{maQRDecompositionBasedMatrix2011}
  L.~Ma, K.~Dickson, J.~McAllister, and J.~McCanny, ``{{QR decomposition-based
    matrix inversion}} for {{high performance embedded MIMO receivers}},''
    \emph{IEEE Trans. Signal Process.}, vol.~59, no.~4, pp.~1858--1867, Apr. 2011.
  
  \bibitem{trefethen1997numerical}
  L.~N. Trefethen and D.~Bau~III, \emph{Numerical linear algebra}.\hskip 1em plus
    0.5em minus 0.4em\relax SIAM, 1997, vol.~50.
  
  \bibitem{3gpp38901}
  \BIBentryALTinterwordspacing
  {3GPP TR 38.901}, ``Study on channel model for frequencies from 0.5 to 100
    {GHz} ({Release} 17),'' Tech. Rep., Apr. 2022. [Online]. Available:
    \url{https://www.etsi.org/deliver/etsi_tr/138900_138999/138901/17.00.00_60/tr_138901v170000p.pdf}
  \BIBentrySTDinterwordspacing
  
  \bibitem{cespedesProbabilisticMIMOSymbol2018}
  J.~C{\'e}spedes, P.~M. Olmos, M.~{S{\'a}nchez-Fern{\'a}ndez}, and
    F.~{Perez-Cruz}, ``Probabilistic {{MIMO symbol detection with expectation
    consistency approximate inference}},'' \emph{IEEE Trans. Veh.
    Technol.}, vol.~67, no.~4, pp. 3481--3494, Apr. 2018.
  
  \bibitem{jaeckel2014quadriga}
  S.~Jaeckel, L.~Raschkowski, K.~Börner, and L.~Thiele, ``{QuaDRiGa}: A 3-D
    multi-cell channel model with time evolution for enabling virtual field
    trials,'' \emph{IEEE Trans. Antennas Propag.}, vol.~62,
    no.~6, pp. 3242--3256, Jun. 2014.
  
  {\bibitem{FundamentalsStatisticalSignal}
  S.~M. Kay, \emph{Fundamentals of Statistical Signal Processing: Estimation
  Theory}.\hskip 1em plus
  0.5em minus 0.4em\relax Englewood Cliffs, NJ, USA: Prentice-Hall, 1993.}

  {\bibitem{weberImperfectChannelstateInformation2006}
  T.~Weber, A.~Sklavos, and M. Meurer, ``Imperfect channel-state information in MIMO transmission,'' \emph{IEEE Trans. Commun.},
		vol 54, no. 3, pp. 543--552, Mar. 2006.}
  
  \end{thebibliography}


\end{document}